\documentclass[prb,twocolumn,showpacs,floatfix,linenumbers]{revtex4-1}
\usepackage{graphicx,color}
\usepackage{amssymb}
\usepackage{amsmath}
\usepackage{xspace}
\usepackage{url}
\usepackage{lineno}
\usepackage{subfigure}

\newcommand{\pdagger}{{\phantom{\dagger}}}
\let \k \relax
\newcommand{\k}{{\bf k}}
\newcommand{\q}{{\bf q}}
\newcommand{\Q}{{\bf Q}}

\newcommand{\beps}{\bar\varepsilon}

\begin{document} 
\title{Chiral charge order in 1$T$-TiSe$_2$: importance of lattice degrees of freedom}
%
\author{B. Zenker$^1$, H. Fehske$^1$, H. Beck$^2$, C. Monney$^3$, and A. R. Bishop$^4$}
\affiliation{$^1$Institut f{\"u}r Physik,
                  Ernst-Moritz-Arndt-Universit{\"a}t Greifswald,
                  D-17489 Greifswald, Germany \\
                  $^2$D{\'e}partement de Physique and Fribourg Center for Nanomaterials, 
                  Universit{\'e} de Fribourg,
                  CH-1700 Fribourg, Switzerland \\
                  $^3$Fritz-Haber-Institut der Max Planck Gesellschaft, 
                  Faradayweg 4-6, 
                  14195 Berlin, Germany \\
                  $^4$Theory, Simulation, and Computation Directorate,
                  Los Alamos National Laboratory, 
                  Los Alamos, New Mexico 87545, USA}

\date{\today}
\begin{abstract}
We address the question of the origin of the recently discovered chiral property of the charge-density-wave phase in 1$T$-TiSe$_2$
which so far lacks a microscopic understanding. We argue that the lattice degrees of freedom seems to  be crucial for this novel phenomenon.
We motivate a theoretical model that takes into account one valence and three conduction bands, a strongly screened Coulomb interaction between the electrons, as well as the coupling of the electrons to a transverse optical phonon mode. The Falicov-Kimball model extended in this way possesses a charge-density-wave state at low temperatures, which is accompanied by a periodic lattice distortion. The charge ordering is driven by a lattice  deformation and electron-hole pairing (excitonic) instability in combination. 
We show  that both electron-phonon interaction and phonon-phonon interaction must be taken into account at least up to quartic order in the lattice displacement to achieve a stable chiral charge order. 
The chiral property is exhibited in the ionic displacements. Furthermore, we provide the ground-state phase diagram of the model and give an estimate of the electron-electron and electron-phonon interaction constants for 1$T$-TiSe$_2$.  
\end{abstract}
\maketitle

\section{Motivation} 
Charge-density-waves (CDWs) brought about by electron-phonon\cite{Pe55} or electron-electron\cite{So79} interactions are broken-symmetry ground states, typically of low-dimensional (D) solids with anisotropic properties.\cite{GR00}  A prominent material exhibiting such a  periodic real-space modulation of its charge density is the transition-metal dichalcogenide 1$T$-TiSe$_2$. This quasi-2D system undergoes a structural phase transition at about $200$ K, at which a commensurate $2\times 2\times 2$ superstructure accompanied by a CDW 
develops.\cite{SMW76} Thereby the CDW features three coexisting components and, for this reason, is denoted as triple CDW. Although the charge-ordered phase in 1$T$-TiSe$_2$ has been a matter of intensive research for more than three decades, the driving force behind the phase transition has not been identified conclusively.

\begin{figure}[b]
\centering
\subfigure[\, nonchiral]{\includegraphics[width=0.3\linewidth]{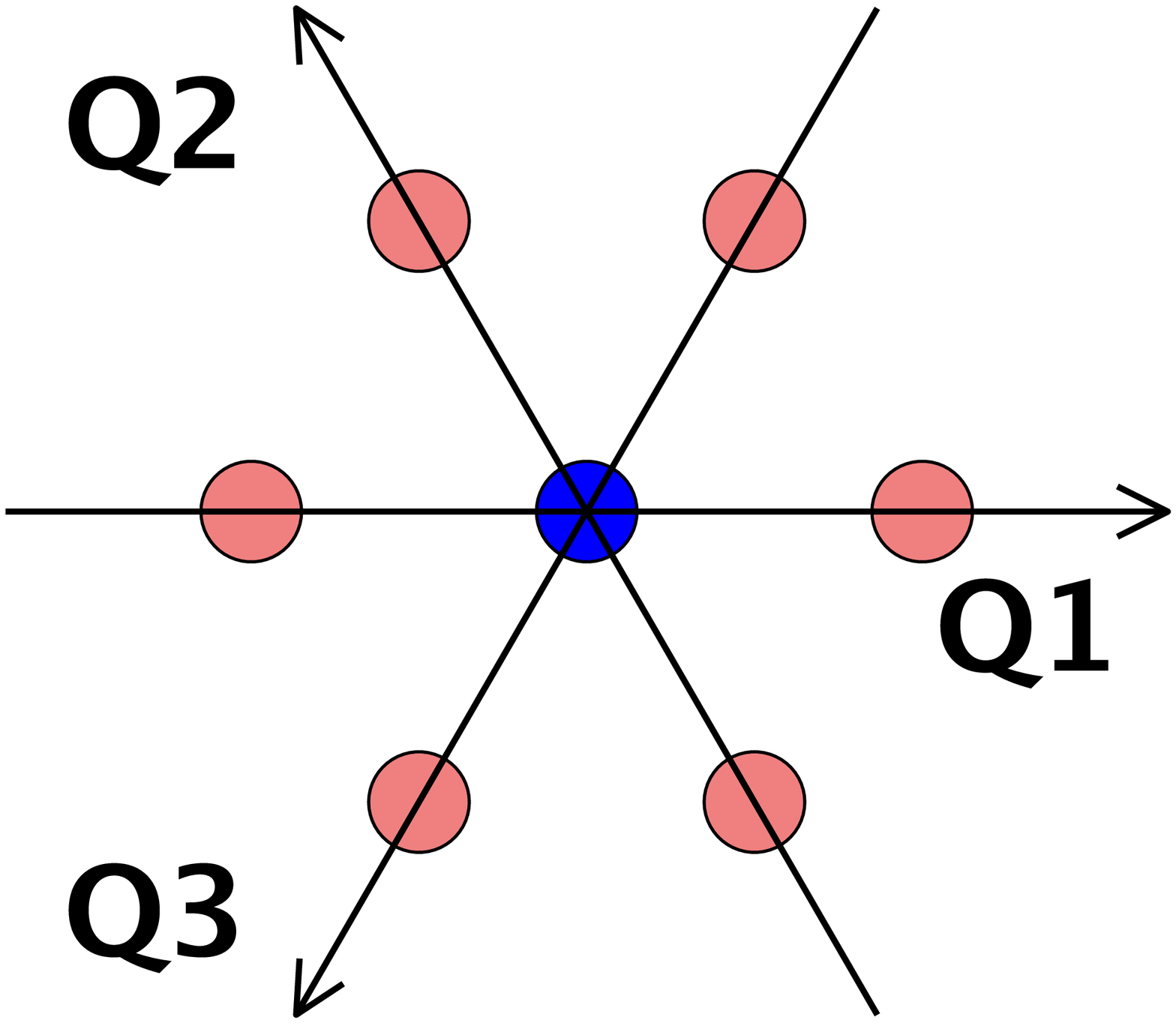}}
\subfigure[\, clockwise]{\includegraphics[width=0.3\linewidth]{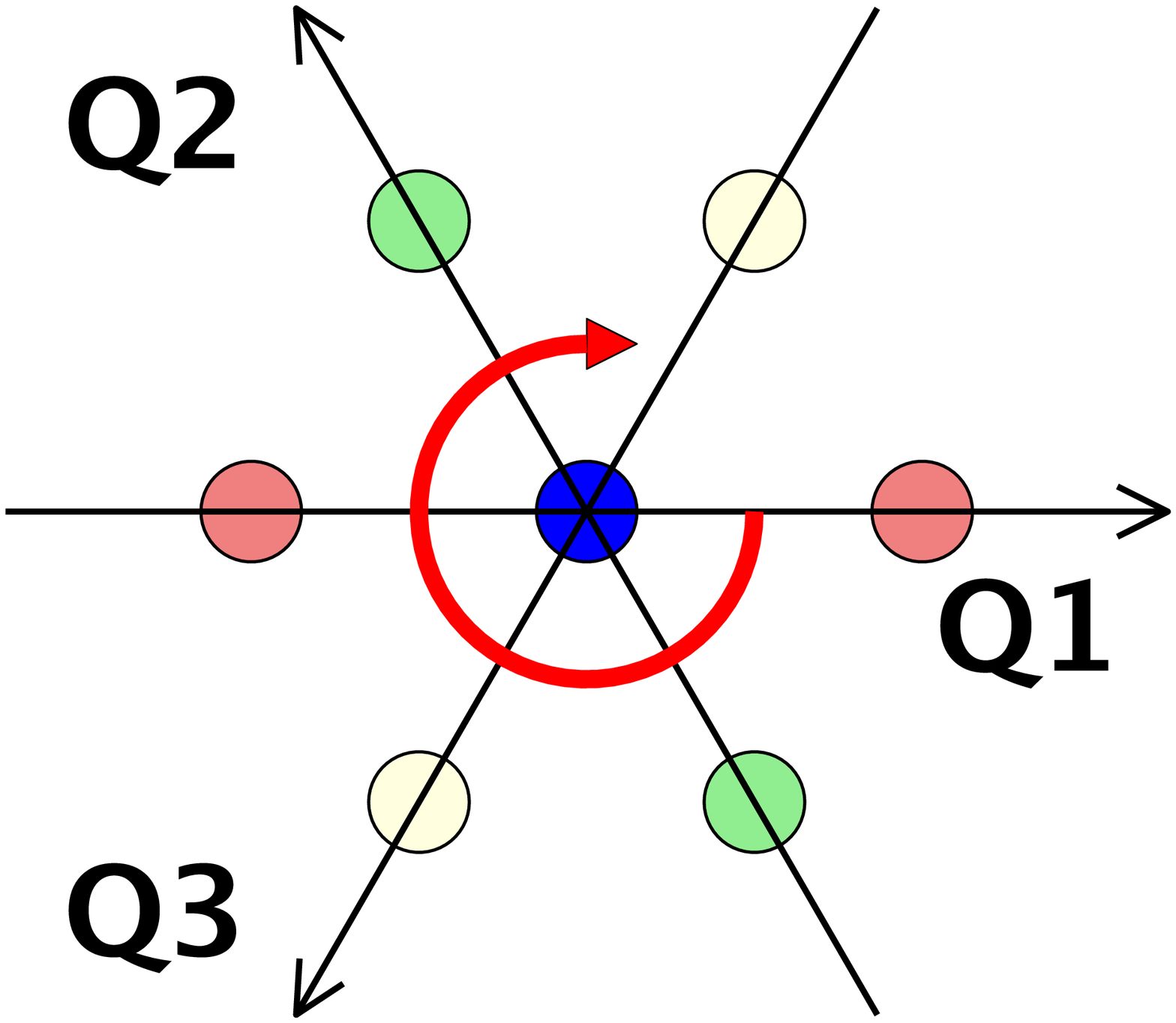}}
\subfigure[\, anticlockwise ]{\includegraphics[width=0.3\linewidth]{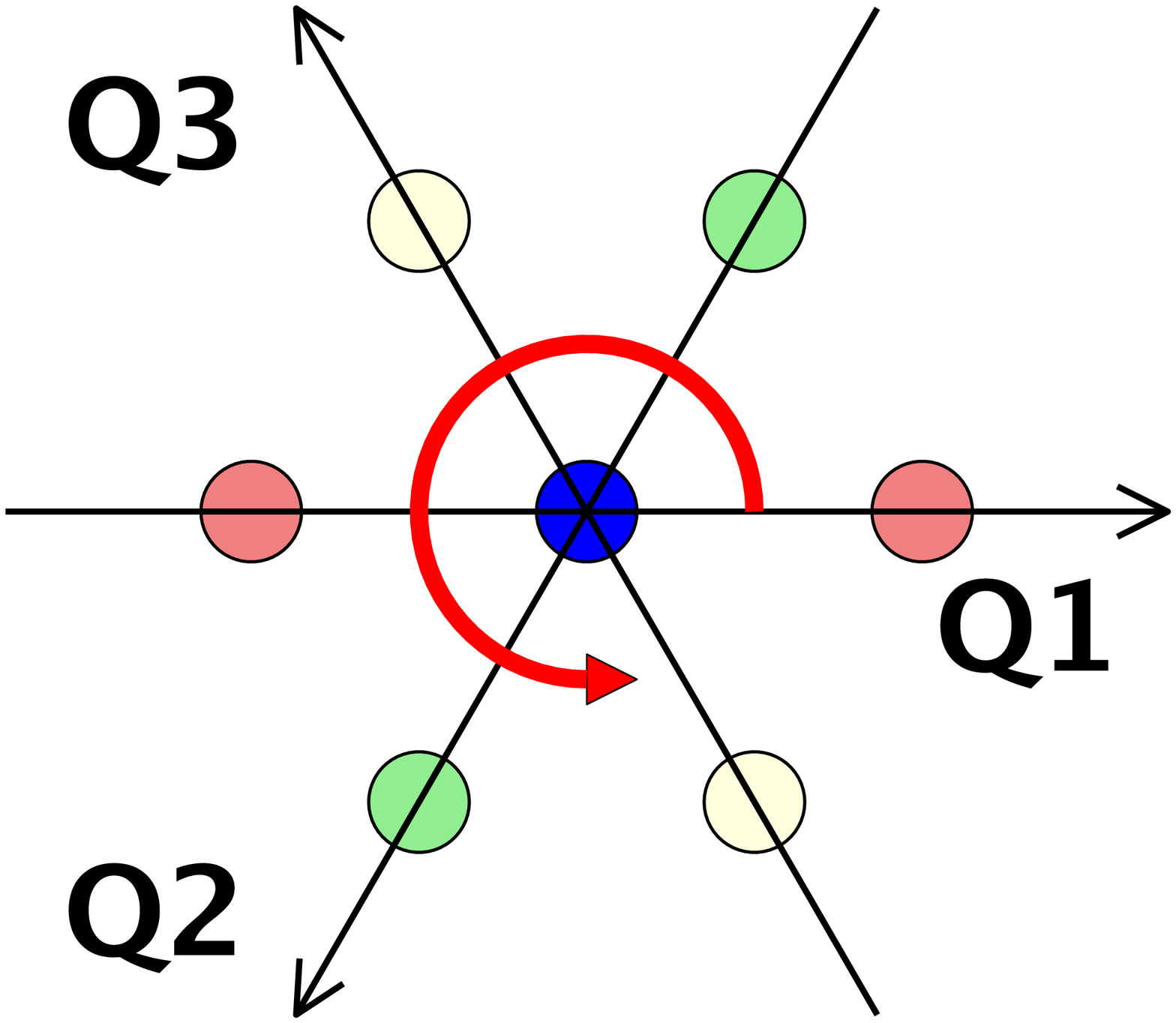}}
\caption{(color online) Electron-density pattern for a triangular lattice in case of (a) a nonchiral CDW or (b,c) chiral CDWs. Filled circles picture the value of the charge densities, where equal colors mark equal densities. For the nonchiral CDW shown in (a) the density modulation along the ordering vectors $\Q_1$, $\Q_2$, and $\Q_3$ is equal. Reflection along an ordering vector yields the same density pattern, i.e., mirror symmetry exists. The situation changes for a chiral CDW. A clockwise   CDW (red arrow) is illustrated in (b). Now  reflexion along $\Q_1$ yields the situation depicted in panel (c). Obviously the pattern (c) describes an anticlockwise CDW: That is, for a chiral CDW mirror symmetry is broken.}
\label{fig:mirror_sym}
\end{figure}

Recent experiments on 1$T$-TiSe$_2$, pointing to a very unusual chiral property of the CDW, have reinforced the interest in this 
problem.\cite{ILSKITOT10,WL10} An object exhibits chirality if it cannot be mapped on its mirror image solely by rotations and translations. 
For a CDW phase characterized by a scalar quantity such  chirality has not been detected before. 
The scanning tunneling microscopy  measurements performed by Ishioka and co-workers, however, show that the amplitude of the tunneling current modulates differently along the CDW unit vectors in 1$T$-TiSe$_2$.\cite{ILSKITOT10}  Since the tunneling-current amplitude directly measures the local electron density, the  charge density modulates differently along the three unit vectors. As a result  the material in its low-temperature phase will not exhibit a three-fold symmetry as suggested by the triangular lattice structure.  The Fourier transformation of the scanning tunneling microscopy  data demonstrates a triple CDW as well as a different charge modulation along each CDW component with the respective ordering vector $Q_\alpha$, $\alpha=1,2,3$.\cite{ILSKITOT10} If one orders the triple-CDW components according to their charge modulation amplitude in ascending order, in a sense a direction is singled out and the triple CDW exhibits chirality because the mirror symmetry is broken, in contrast to usual CDWs;\cite{WL10} see the schematic representation by Fig.~\ref{fig:mirror_sym}.
Note that clockwise and anticlockwise chiral CDWs were found in the same sample, suggesting that these states are degenerate. This two-fold symmetry is corroborated by optical polarimetry measurements.\cite{ILSKITOT10} Ishioka and co-workers furthermore noticed that the experimental data can be reproduced by a charge density modulation of the form 
\begin{equation}
n_i(\Q_\alpha) = A \cos(\Q_\alpha {\bf R}_i + \theta_\alpha),
\label{assume_locDens}
\end{equation}
where $A$ is the modulation amplitude and $\theta_\alpha$ are initial phases.\cite{ILSKITOT10}  For a chiral CDW to exist the phases of the CDW components must differ, i.e., $\theta_1\neq \theta_2\neq \theta_3$.

From a theoretical point of view the chiral CDW in 1$T$-TiSe$_2$ was addressed by a Landau-Ginzburg approach.\cite{We11,We12} Thereby the relative phases of the CDW order parameters were obtained by minimizing the free energy functional. Two CDW transitions were found with decreasing  temperature: Firstly a standard (nonchiral) CDW appears, and subsequently a chiral CDW emerges, i.e., $T_{\rm nonchiral\, CDW} > T_{\rm chiral \,CDW}$. Within the CDW phase three distinct orbital sectors are occupied, leading to an orbital-ordered state and three interacting lattice displacement waves (with different polarizations).  

An open issue is the microscopic mechanism driving the CDW transition. Basically two scenarios have been discussed in the literature, 
where the charge order results from purely electronic, respectively electron-lattice, correlations. Angle-resolved photoemission spectroscopy  data reveal a relatively large transfer of spectral weight from the original bands to the back-folded bands (due to the CDW transition), compared with the small ionic displacement. This suggests an electronic mechanism within the excitonic insulator (EI) scenario.\cite{CMCBDGBA07, MCCBSDGBA09} A corresponding tight-binding calculation estimates the amplitude of the lattice deformation caused by an EI instability to be of the same order as the measured one.\cite{MBCAB11} The gradual 
suppression of the CDW phase by changing solely electronic properties by intercalation with S or Te further corroborates the EI concept.\cite{MBJM11} Most convincingly, time-resolved photoemission spectroscopy reveals an extremely fast response of the CDW to external light pulses, which favors an electronic mechanism.\cite{RHWSSSCLAKMKRB11} Alternatively, the coupling to the lattice degrees of freedom may drive the CDW transition, e.g.,  by a cooperative Jahn-Teller effect.\cite{Hu77,RKS02}  Here  the particular form of the phonon dispersion and the softening of transverse optical phonon modes were elaborated within a tight-binding approach and found to agree with
the experimental results.\cite{YM80, MSYT81, SYM85, HZHCC01} The same holds for an {\it ab-initio} approach~\cite{WRCOKHHBA11} to a Jahn-Teller effect. Likewise the onset of superconductivity by applying pressure may be understood within a phonon-driven CDW scenario.\cite {CM11} Since some properties of the CDW in 1$T$-TiSe$_2$  can be understood by the excitonic condensation 
of electron-hole pairs and others by the instability of a phonon mode, a combined scenario has been proposed.\cite{WNS10}

As yet it is unclear whether the chiral property of the CDW favors the electronic or lattice scenario, or a combination of both.
In the present work, this issue is addressed amongst others. We start by investigating the CDW from an EI perspective. 
The corresponding mean-field approach for an extended Falicov-Kimball model is presented in Sec.~\ref{sec:ElectronicDegrees}. We show that the EI scenario is insufficient to explain a stable chiral CDW. We proceed by including the lattice degrees of freedom. 
We find that the electron-phonon interaction and the phonon-phonon interaction both must be taken into account at least up to quartic order in the lattice distortion in order to stabilize chiral charge order.
This is elaborated in Sec.~\ref{sec:LatticeDegrees},  
and in Sec.~\ref{sec:GS_energy} we present the ground-state energy as a function of the static lattice distortion. In Sec.~\ref{sec:Ph_boundary} the CDW phase boundary is derived.
The CDW state is characterized analytically in Sec.~\ref{sec:PhTransitionChar}. Section~\ref{sec:NumericalResults} contains our numerical results. Here we give the functional dependences of the relevant phases on the lattice distortion, show the finite-temperature phase diagram, derive the ground-state phase diagram, and estimate the interaction constants for 1$T$-TiSe$_2$. In Sec.~\ref{sec:Summary}, we summarize and conclude.

\section{Model and theoretical approach}

\subsection{Electronic degrees of freedom}
\label{sec:ElectronicDegrees}

\subsubsection{Band structure}
\begin{figure}[h]
\centering
\includegraphics[width=0.49\linewidth]{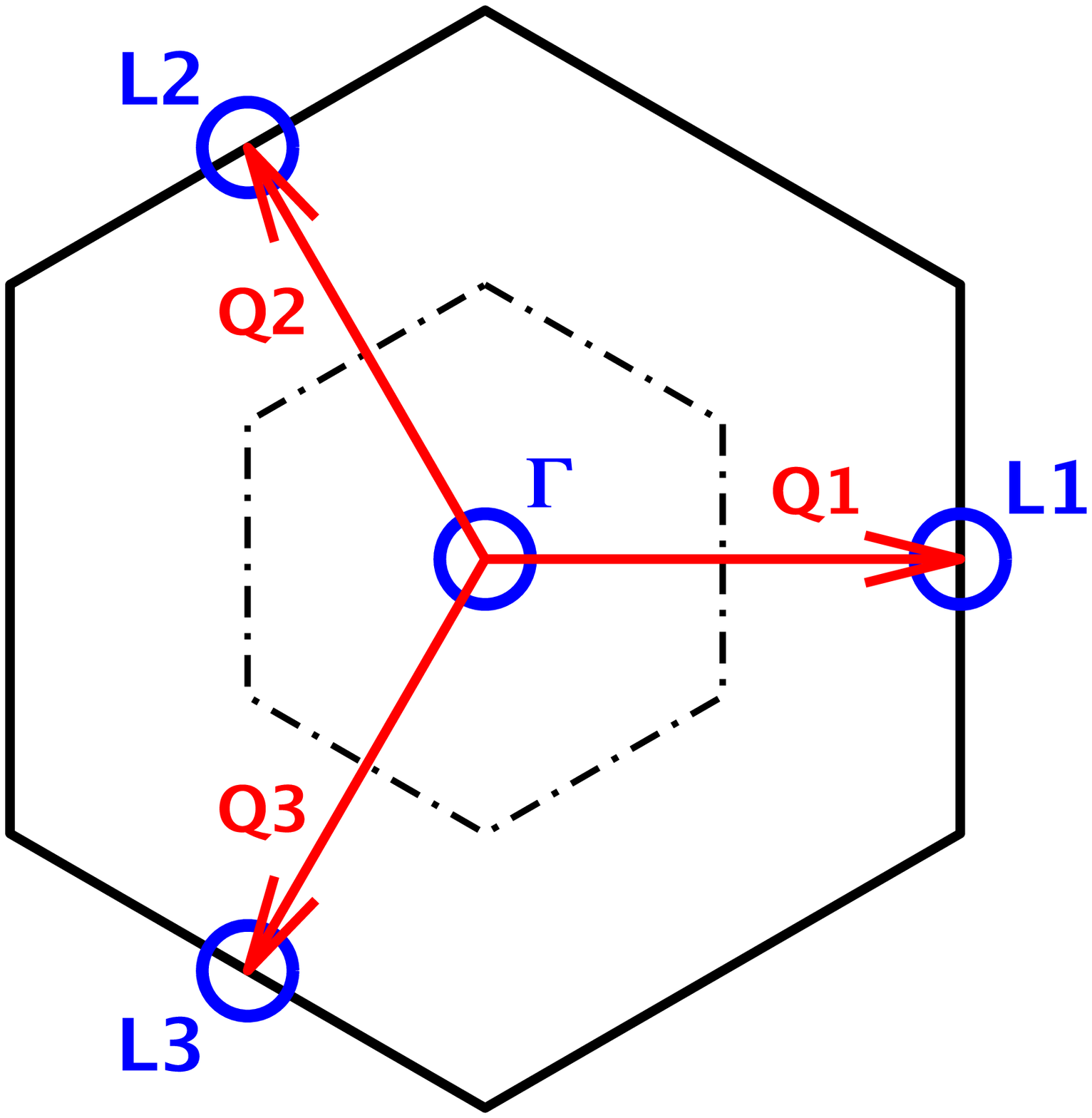}
\includegraphics[width=0.49\linewidth]{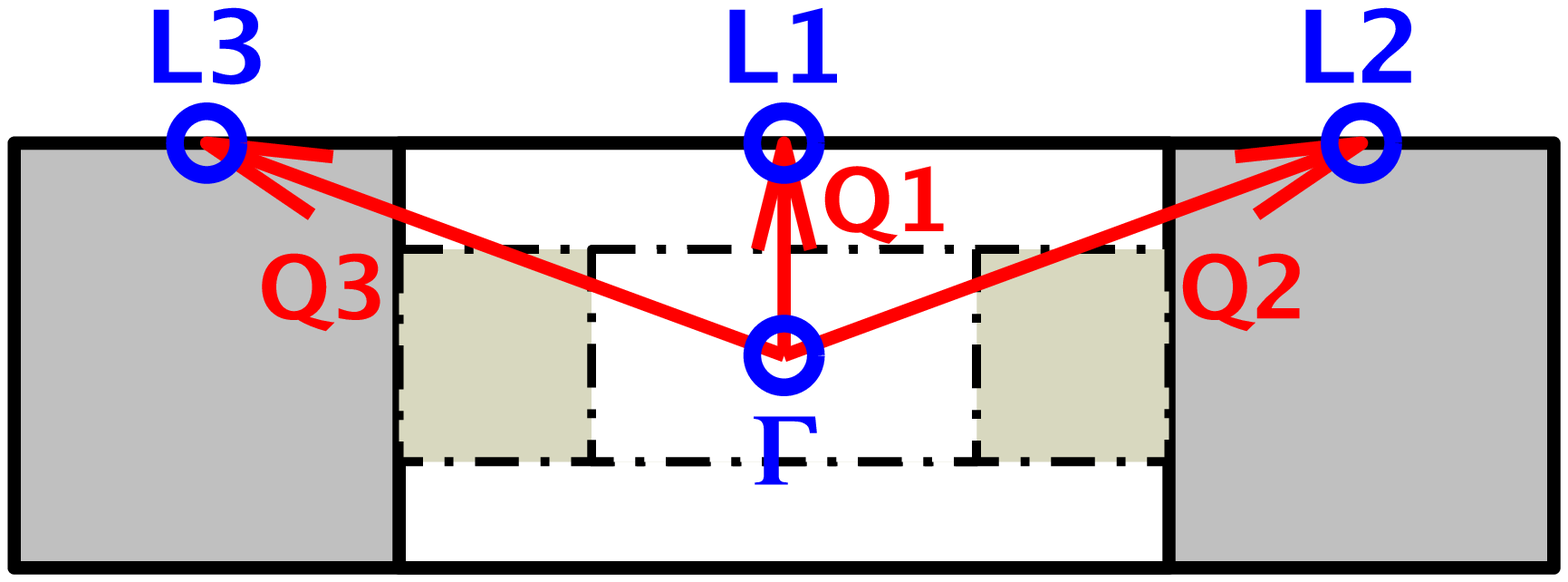}
\caption{(color online) First Brillouin zone (BZ) of 1$T$-TiSe$_2$ with high-symmetry points in the normal phase (solid line) and in the CDW phase (dot-dashed line). Red arrows show the CDW ordering vectors. Left panel: projection onto the $xy$-plane, right panel: 
projection onto the $yz$-plane.}
\label{fig:BZ}
\end{figure}

\begin{figure}[h]
\centering
\subfigure[band structure along high-symmetry directions of the~BZ]{\includegraphics[height=0.41\linewidth]{fig3a_all_Bands_line.eps}}
\subfigure[band dispersion close to the Fermi level]{\includegraphics[height=0.42\linewidth]{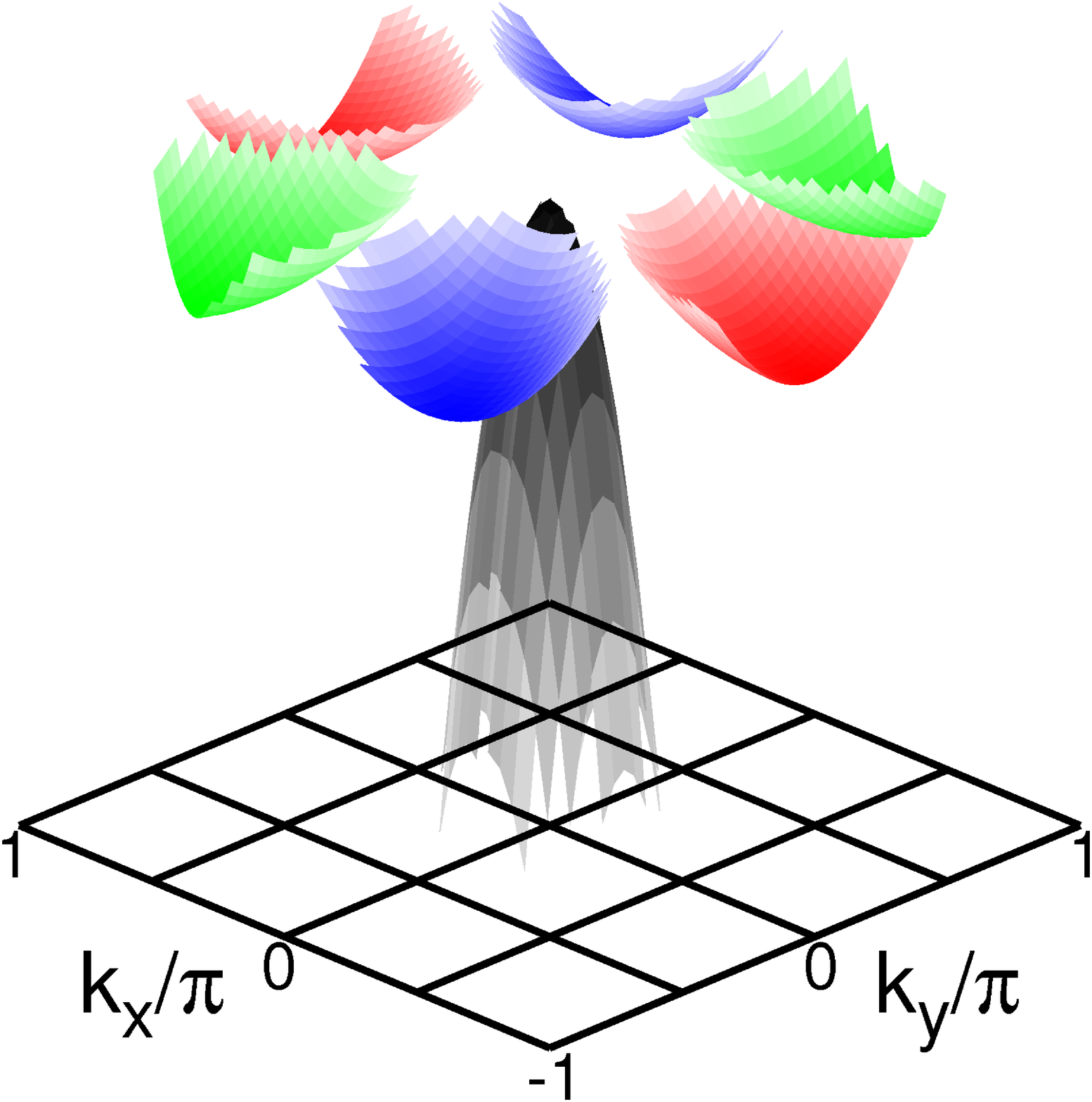}}
\caption{(color online)  Model band structure in the normal phase. The valence band is colored  black, and the conduction bands are colored red, blue, and green.}
\label{fig:bands}
\end{figure}

Since the electronic properties of 1$T$-TiSe$_2$ are dominated by the electrons near the Fermi energy, in what follows  we  take into account only the top valence band and the lower conduction band. The maximum of the valence band is located at the $\Gamma$-point. 
The conduction band exhibits  minima at the three $L$-points, see Fig.~\ref{fig:BZ}. 
To facilitate the notation, we artificially split the conduction band into three symmetry-equivalent bands indexed by $\alpha$, each having one minimum at the point $L_\alpha$. The band dispersions of these three conduction bands mimics the true band structure 
close to the $L$-points.\cite{MCCBSDGBA09} Figure~\ref{fig:bands} illustrates the situation close to the Fermi level.
Then the  free electron part is written as
\begin{equation}
H_{\rm e} = \sum_{\k} \varepsilon_{\k f} f_\k^\dagger f_\k^\pdagger + \sum_{\k,\alpha} \varepsilon_{\k \alpha} c_{\k \alpha}^\dagger c_{\k \alpha}^\pdagger ,
\label{H_e}
\end{equation}
where $f_\k^{(\dagger)}$ annihilates (creates) an electron in the valence band with momentum $\k$ and $c_{\k \alpha}^{(\dagger)}$ annihilates (creates) an electron in the conduction band with momentum $\k$ and band index $\alpha$. The corresponding valence-band dispersion and the conduction-band dispersions are denoted as $\varepsilon_{\k f}$ and $ \varepsilon_{\k \alpha}$. They will be specified in Sec.~\ref{sec:mod_params}. The spin of the electrons is neglected.

Taking the band structure and a band filling factor $n=1/4$ into account, 1$T$-TiSe$_2$ resides in the vicinity of  a semimetal-semiconductor transition, cf. Fig~\ref{fig:bands}. Accordingly  the chemical potential $\mu$ is determined by 
\begin{equation}
n_f+\sum_\alpha n_\alpha=1,
\label{densConst}
\end{equation}
where
\begin{eqnarray}
n_f &=& \frac{1}{N} \sum_\k \langle n_\k^f \rangle = \frac{1}{N} \sum_\k \langle f_\k^\dagger f_\k^\pdagger \rangle , \\
n_\alpha &=& \frac{1}{N} \sum_\k \langle n_\k^\alpha\rangle = \frac{1}{N} \sum_\k \langle c_{\k\alpha}^\dagger c_{\k\alpha}^\pdagger \rangle\,. 
\end{eqnarray}
Here  $N$ denotes the total number of lattice sites. 

Regarding  the isotropy (anisotropy)  of the valence (conduction) band(s) the Fermi surface of 1$T$-TiSe$_2$ is only poorly  nested,\cite{RKS02} 
which rules out a nesting mechanism  for the CDW formation even in a simplified 2D setting.

\subsubsection{Electron-electron interaction}
Due to the strong screening of the Coulomb interaction in 1$T$-TiSe$_2$,\cite{MMAB12b} we assume a local electron-electron interaction,
\begin{eqnarray}
H_{\rm e-e} &=& \frac{U_{cc}}{N} {\sum_{\k,\k',\q}\sum_{\alpha}\sum_{\beta>\alpha}} c_{\k+\q \alpha}^\dagger c_{\k\alpha}^\pdagger c_{\k'\beta}^\dagger c_{\k'+\q\beta}^\pdagger\nonumber \\
&&+\frac{U_{fc}}{N}\sum_{\k,\k',\q}\sum_{\alpha} f_{\k+\q}^\dagger f_{\k}^\pdagger c_{\k' \alpha}^\dagger c_{\k'+\q \alpha}^\pdagger\,,
\label{H_ee}
\end{eqnarray}
where  $U_{cc}$ denotes the Coulomb repulsion among the conduction electrons. The on-site Coulomb interaction $U_{fc}$ 
between valence and conduction band electrons determines the distribution of electrons between  these ``subsystems'' and therefore may drive a valence transition, as observed, e.g., in heavy fermion and intermediate-valence Tm[Se,Te] compounds.\cite{NW90,WB13} 
If the total electronic model contains an explicit hybridization between $f$ and $c$ electrons~\cite{KMM76,POS96a} or, as in our case, dispersive $c$ {\it and} $f$ bands,\cite{Ba02b} coherence between $c$ and $f$ particles can develop. Then $U_{fc}$ may lead to a
pairing of  $c$-band electrons and $f$-band holes, i.e., to the formation of excitons, and, provided a large enough number of excitons is created, a subsequent spontaneous condensation of these composite quasiparticles may develop. In real systems this excitonic instability is expected to occur, when semimetals with very small band overlap or semiconductors with  very small band gap are cooled to extremely low temperatures.\cite{Mo61,Kno63} The excitonic condensate then typifies a macroscopic phase-coherent insulating state, the EI, which separates the semimetal from the insulator.\cite{JRK67,BF06}  From a theoretical point of view, Falicov-Kimball-type models seem to be the most promising candidates for realizing collective exciton phases. This holds particularly for the generic two-band extended Falicov-Kimball model (EFKM), where an EI ground state has been proven to exist in 1D and 2D by constrained-path Monte Carlo 
simulations.\cite{Ba02b,BGBL04} Subsequent Hartree-Fock, slave-boson and projector-based renormalization techniques yield the 2D EFKM ground-state phase diagram in even quantitative accordance with unbiased Monte Carlo data,\cite{BGBL04,Fa08,SC08,IPBBF08,ZIBF10,PBF10,ZIBF11,PFB11,SEO11,ZIBF12} supporting the applicability of these analytical approaches also in 3D and for more 
complicated situations.

The electronic part of our Hamiltonian, 
\begin{equation}
H_{\rm mEFKM}=H_{\rm e}+H_{\rm e-e} ,
\label{def_Hmefkm}
\end{equation}
can be viewed as a multiband extended Falicov-Kimball model (mEFKM). We note that the mEFKM was studied
previously and has been shown to reproduce the angle-resolved photoemission spectroscopy  data for 1$T$-TiSe$_2$ at temperatures below the critical temperature,\cite{CMCBDGBA07,MCCBSDGBA09, MSGMDCMBTBA10, MSGMDBACMBBT10, CCHMAOO12}  
as well as above but close to the critical temperature.\cite{MMAB12a, MMAB12b} 

We note that the mEFKM exhibits a particular U(1) symmetry. This can be seen by applying the unitary transformation $U_{\varphi,\alpha}=e^{i\varphi S_\alpha}$ with $S_\alpha=\tfrac{1}{2}\sum_i( f_i^\dagger f_i^\pdagger - c_{i \alpha}^\dagger c_{i\alpha}^\pdagger)$. The operators $f_i^{(\dagger)}$ and $c_{i\alpha}^{(\dagger)}$ annihilate (create) an electron at Wannier site $i$. Obviously we have
\begin{equation}
H_{\rm mEFKM}= U_{\varphi,\alpha}\,  H_{\rm mEFKM} \,U_{\varphi,\alpha}^\dagger\,.
\label{trafoH}
\end{equation}
This symmetry leads to a degeneracy between chiral and nonchiral CDWs (see below).

To proceed, we perform a Hartree-Fock  decoupling of  the electron-electron interaction terms:
\begin{align}
&\frac{U_{cc}}{N}{\sum_{\k,\k',\q}\sum_{\alpha}\sum_{\beta>\alpha}}\, c_{\k+\q\alpha}^\dagger  c_{\k\alpha}^\pdagger c_{\k'\beta}^\dagger c_{\k'+\q\beta}^\pdagger \nonumber \rightarrow \label{decoup1}\\
&\qquad U_{cc} \sum_\k \sum_\alpha \sum_{\beta\neq\alpha} c_{\k\alpha}^\dagger c_{\k\alpha}^\pdagger n_\beta 
- NU_{cc} \sum_\alpha \sum_{\beta>\alpha} n_\alpha n_\beta\,,\\
&\frac{U_{fc}}{N}\sum_{\k,\k',\q}\sum_{\alpha} f_{\k+\q}^\dagger  f_\k^\pdagger c_{\k'\alpha}^\dagger c_{\k'+\q\alpha}^\pdagger \nonumber \rightarrow\\
&\qquad U_{fc} \sum_\alpha  n_{\alpha} \sum_\k f_\k^\dagger f_\k^\pdagger + U_{fc} n_f \sum_{\k,\alpha} c_{\k\alpha}^\dagger c_{\k\alpha}^\pdagger 
 \nonumber \\
&\qquad- NU_{fc} n_f\sum_\alpha n_\alpha-\sum_\alpha \Delta_{\Q\alpha} \sum_\k c_{\k+\Q_\alpha \alpha}^\dagger f_\k^\pdagger \nonumber \\
&\qquad- \sum_\alpha \Delta_{\Q\alpha}^\ast \sum_\k f_\k^\dagger c_{\k+\Q_\alpha \alpha}^\pdagger
+\frac{N}{U_{fc}} \sum_\alpha |\Delta_{\Q\alpha}|^2\,. 
\label{decoup2}
\end{align}
Here we introduced the EI order parameter functions
\begin{eqnarray}
\Delta_{\Q\alpha} &=& \frac{U_{fc}}{N} \sum_\k \langle f_\k^\dagger c_{\k+\Q_\alpha \alpha}^\pdagger \rangle ,\\
\Delta_{\Q\alpha}^\ast &=& \frac{U_{fc}}{N} \sum_\k \langle c_{\k+\Q_\alpha \alpha}^\dagger f_\k^\pdagger \rangle . \label{EIOP_def}
\end{eqnarray}
Since the experiments on 1$T$-TiSe$_2$ suggest that the spontaneous hybridization of the valence band with one of the three conduction bands is the dominant  effect of the electron-electron interaction,\cite{CMCBDGBA07} in deriving Eq.~\eqref{decoup1}, we neglected all terms that mix different conduction bands. 

The resulting decoupled Hamiltonian takes the form
\begin{eqnarray}
\bar H_{\rm mEFKM} &=& \sum_\k \beps_{\k f} f_\k^\dagger f_\k^\pdagger + \sum_{\k,\alpha} \beps_{\k\alpha} c_{\k\alpha}^\dagger c_{\k\alpha}^\pdagger
\nonumber \\
&&-\sum_{\k,\alpha} \Delta_{\Q \alpha} c_{\k+\Q_\alpha \alpha}^\dagger f_\k^\pdagger 
- \sum_{\k,\alpha} \Delta_{\Q\alpha}^\ast f_\k^\dagger c_{\k+\Q_\alpha \alpha}^\pdagger
\nonumber \\
&&- NU_{fc} n_f \sum_\alpha n_\alpha - NU_{cc} \sum_\alpha \sum_{\beta>\alpha} n_\alpha n_\beta 
\nonumber \\
&& +\frac{N}{U_{fc}}\sum_\alpha |\Delta_{\Q\alpha}|^2 \,,
\label{H_dec}
\end{eqnarray}
with shifted $f$- and $c$-band dispersions:
\begin{eqnarray}
\beps_{\k f}&=&\varepsilon_{\k f} + U_{fc} \sum_\alpha n_\alpha \label{beps_f} , \\
\beps_{\k \alpha}&=&\varepsilon_{\k \alpha} + U_{fc} n_f+U_{cc}\sum_{\beta\neq\alpha} n_\beta \label{beps_c} \,.
\end{eqnarray}

The EI low-temperature phase is characterized by nonvanishing expectation values $\langle f_\k^\dagger c_{\k+\Q_\alpha \alpha}^\pdagger \rangle$, $\langle c_{\k+\Q_\alpha \alpha}^\dagger f_\k^\pdagger \rangle$, which cause a correlation gap in the excitation spectrum. 
The mean local electron density in the EI phase is  
\begin{equation}
n_i= 1+\frac{2}{N}\sum_{\k ,\alpha}  |\langle c_{\k+\Q_\alpha \alpha}^\dagger f_\k^\pdagger \rangle|  \cos(\Q_\alpha{\bf R}_i+\theta_\alpha) ,
\label{locDens}
\end{equation}
where
\begin{equation}
\frac{1}{N}\sum_\k \langle c_{\k+\Q_\alpha \alpha}^\dagger f_\k^\pdagger \rangle
 = \frac{1}{N}\sum_\k |\langle c_{\k+\Q_\alpha \alpha}^\dagger f_\k^\pdagger \rangle| e^{i\theta_\alpha} 
 =\frac{\Delta_{\Q\alpha}^\ast}{U_{fc}} .
\label{EIlikeOP} 
\end{equation}
Comparing Eq.~\eqref{locDens} with relation~\eqref{assume_locDens} we recognize the amplitude of the charge density modulation as the modulus of the hybridization function $\sum_\k \langle c_{\k+\Q_\alpha \alpha}^\dagger f_\k^\pdagger \rangle$. Likewise we can identify the initial phases $\theta_\alpha$ in the density modulation as the phases of the hybridization functions (which coincide with the phases of the EI order parameters). 

Note that previous theoretical studies of the mEFKM\cite{MCCBSDGBA09,MBCAB11} did not include the phase differences of the $\theta_\alpha$, which will be essential for the establishment of a chiral CDW.\cite{ILSKITOT10, WL10, We11, We12} 
If one is not concerned with the chiral CDW problem,  disregarding the phases $\theta_\alpha$ seems to be justified since 
the U(1) symmetry of the mEFKM prevents the appearance of a stable chiral CDW anyway. We show this by analyzing the behavior of the electron operators under the unitary transformation $U_{\varphi,\alpha}$: 
$\tilde c_{i \alpha}^{(\dagger)}=U_{\varphi,\alpha} c_{i \alpha}^{(\dagger)} U_{\varphi,\alpha}^\dagger$ 
and $\tilde f_{i}^{(\dagger)}=U_{\varphi,\alpha} f_{i}^{(\dagger)} U_{\varphi,\alpha}^\dagger$.
The hybridization functions (in real space) then transform as $\langle c_{i\alpha}^\dagger f_i\rangle e^{i\Q_\alpha {\bf R}_i}=e^{-i\varphi} \langle \tilde c_{i\alpha}^\dagger \tilde f_i\rangle e^{i\Q_\alpha {\bf R}_i}$. That is, the phases $\theta_\alpha$ can be controlled by the unitary transformation through the angles $\varphi$.  However, in view of~\eqref{trafoH} the total energy is independent of the $\theta_\alpha$. 
Hence these phases can be chosen arbitrarily, and there is no mechanism that stabilizes a given phase difference. Therefore the mEFKM is insufficient to describe a chiral CDW in 1$T$-TiSe$_2$. In the following we will demonstrate that the coupling of the electrons to the lattice degrees of freedom can break the U(1) symmetry of the mEFKM  and consequently can stabilize a chiral CDW.

\subsection{Lattice degrees of freedom}
\label{sec:LatticeDegrees}

\subsubsection{Electron-phonon coupling}
For 1$T$-TiSe$_2$ there are experimental and theoretical evidences that the weak periodic lattice distortion observed comes from a softening of a transverse optical phonon mode.~\cite{YM80, MSYT81, SYM85, HZHCC01, WRCOKHHBA11} We therefore include a single-mode electron-phonon interaction in our model. 
If we expand the electron-lattice interaction up to quartic order in the lattice distortion, we obtain the electron-phonon interaction as
\begin{equation}
H_{\rm e-ph} = H_{\rm e-ph}^{(1)}+H_{\rm e-ph}^{(2)}+H_{\rm e-ph}^{(3)}+H_{\rm e-ph}^{(4)} ,
\label{H_eph_tot}
\end{equation}
where
\begin{eqnarray}
H_{\rm e-ph}^{(1)} &=& \frac{1}{\sqrt{N}} \sum_{\k,\q} \sum_{\lambda,\lambda'} 
g_1(\k,\q,\lambda,\lambda') 
(b_\q^\dagger + b_{-\q}^\pdagger) c_{\k\lambda}^\dagger c_{\k+\q \lambda'}^\pdagger ,
\nonumber \\
\label{H_eph_1} 
\end{eqnarray}
\begin{eqnarray}
H_{\rm e-ph}^{(2)} &=& \frac{1}{2N} \sum_{\k,\q_1,\q_2} \sum_{\lambda,\lambda'}
 g_2(\k,\q_1,\q_2,\lambda,\lambda')(b_{\q_1}^\dagger + b_{-\q_1}^\pdagger) 
\nonumber \\
&& \times (b_{\q_2}^\dagger + b_{-\q_2}^\pdagger)
c_{\k\lambda}^\dagger c_{\k+\q_1+\q_2 \lambda'}^\pdagger ,
\label{H_eph_2}
\\
H_{\rm e-ph}^{(3)}&=&\frac{1}{6N^{\frac{3}{2}}}\sum_{\k,\q_1,\q_2,\q_3}\sum_{\lambda,\lambda'}
g_3(\k,\q_1,\q_2,\q_3,\lambda,\lambda') 
\nonumber \\
&&\times (b_{\q_1}^\dagger + b_{-\q_1}^\pdagger)  (b_{\q_2}^\dagger + b_{-\q_2}^\pdagger) 
(b_{\q_3}^\dagger + b_{-\q_3}^\pdagger)
\nonumber \\
&&\times 
c_{\k\lambda}^\dagger c_{\k+\q_1+\q_2+\q_3 \lambda'}^\pdagger ,
\label{H_eph_3}
\\
H_{\rm e-ph}^{(4)}&=&\frac{1}{24N^2}\sum_{\k,\q_1,\q_2,\q_3,\q_4}\sum_{\lambda,\lambda'}
g_4(\k,\q_1,\q_2,\q_3,\q_4,\lambda,\lambda') 
\nonumber \\
&&\times (b_{\q_1}^\dagger + b_{-\q_1}^\pdagger) (b_{\q_2}^\dagger + b_{-\q_2}^\pdagger) 
(b_{\q_3}^\dagger + b_{-\q_3}^\pdagger)  
\nonumber \\
&&\times (b_{\q_4}^\dagger + b_{-\q_4}^\pdagger) c_{\k\lambda}^\dagger
 c_{\k+\q_1+\q_2+\q_3+\q_4 \lambda'}^\pdagger ,
\label{H_eph_4}
\end{eqnarray}
where $b_\q^{(\dagger)}$ describes the annihilation (creation) operator of a phonon carrying the momentum $\q$, $g_i$ ($i=1,2,3,4$) denote the electron-phonon coupling constants, and $\lambda$, $\lambda'$ label the band degree of freedom. Most notably the band-mixing terms; i.e., if $\lambda\neq\lambda'$ in Eqs.~\eqref{H_eph_1}-\eqref{H_eph_4}, break the U(1) symmetry of the mEFKM, i.e., the arbitrariness with respect to the phases $\theta_\alpha$ is eliminated.

\subsubsection{Phonon-phonon interaction}
Within the harmonic approximation, the Hamiltonian for the (noninteracting) phonons reads\cite{Zi60}
\begin{equation}
H_{\rm ph}=\sum_{\q} \hbar \omega(\q) b_{\q}^\dagger b_{\q}^\pdagger \,,
\label{H_ph_1}
\end{equation}
where $\omega(\q)$ is the bare phonon frequency. A coupling between the lattice vibrations results from the anharmonic 
contributions in the expansion of the potential for the ions.\cite{AM76} As we will see below, such an explicit phonon-phonon interaction may stabilize the chiral CDW phase. 
We expand the phonon-phonon interaction also up to quartic order in the lattice displacement.
We obtain 
\begin{eqnarray}
H_{\rm ph-ph} &=&  \frac{1}{\sqrt{N}} \sum_{\q_1,\q_2,\q_3 } B(\q_1,\q_2,\q_3) 
( b_{\q_1}^\dagger + b_{-\q_1}^\pdagger) 
\nonumber \\
&&\times ( b_{\q_2}^\dagger + b_{-\q_2}^\pdagger)
( b_{\q_3}^\dagger + b_{-\q_3}^\pdagger)
\nonumber \\
&& +\frac{1}{N} \sum_{\q_1,\q_2,\q_3, \q_4 } D(\q_1,\q_2,\q_3,\q_4) 
( b_{\q_1}^\dagger + b_{-\q_1}^\pdagger) 
\nonumber \\
&&\times ( b_{\q_2}^\dagger + b_{-\q_2}^\pdagger)
( b_{\q_3}^\dagger + b_{-\q_3}^\pdagger)
( b_{\q_4}^\dagger + b_{-\q_4}^\pdagger)\,. 
\end{eqnarray}
The explicit expressions of $B (\q_1, \q_2, \q_3)$ and $D(\q_1,\q_2,\q_3,\q_4)$ are lengthy. We 
note only  the symmetry relations 
\begin{eqnarray}
B(-\q_1,-\q_2,-\q_3)&= &B^\ast(\q_1,\q_2,\q_3)\,,\\ D(-\q_1,-\q_2,-\q_3,-\q_4)&=&D^\ast(\q_1,\q_2,\q_3,\q_4)\,,
 \end{eqnarray}
and point out the constraints 
\begin{eqnarray}
B(\q_1, \q_2, \q_3) &\propto& \delta_{\q_1+\q_2+\q_3,{\bf G}} \,,\label{constraint_B}
\\
D(\q_1,\q_2,\q_3,\q_4) &\propto& \delta_{\q_1+\q_2+\q_3+\q_4,{\bf G}}\,. \label{constraint_D}
\end{eqnarray}
Here ${\bf G}$ is a reciprocal lattice vector of the undistorted lattice.

\subsubsection{Frozen-phonon approach}
We now apply the frozen-phonon approximation and calculate the lattice distortion at low temperatures. As elaborated in Refs.~\onlinecite{MSYT81,SYM85,HZHCC01,WRCOKHHBA11} the phonons causing the lattice displacements in 1$T$-TiSe$_2$ have the momenta $\Q_\alpha$ shown in Fig.~\ref{fig:BZ}. Their softening is inherently connected to strong electronic correlations.~\cite{MMAB12b} 
It has been suggested that the $\Q_1$, $\Q_2$, and $\Q_3$ phonons become soft at the same temperature;~\cite{SYM85} we therefore assume $|g_1(\k,\Q_1,\lambda,\lambda')|=|g_1(\k,\Q_2,\lambda,\lambda')|=|g_1(\k,\Q_3,\lambda,\lambda')|=g_{1\Q}(\k)$, likewise the other electron-phonon coupling constants, and $\omega(\Q_1)=\omega(\Q_2)=\omega(\Q_3)=\omega$. A finite displacement of the ions is characterized by $\langle b_{\Q_\alpha}^\dagger \rangle = \langle b_{-\Q_\alpha} \rangle \neq 0$. We denote the static lattice distortions by
\begin{eqnarray}
\delta_{\Q\alpha} &=& \frac{2}{\sqrt{N}} \langle b_{\Q_\alpha}\rangle 
= |\delta_{\Q_\alpha}|e^{-i\phi_\alpha} ,
\label{def_PhOP1} \\
\delta_{\Q\alpha}^\ast &=& \frac{2}{\sqrt{N}} \langle b_{\Q_\alpha}^\dagger \rangle
= |\delta_{\Q_\alpha}| e^{i\phi_\alpha} .
\label{def_PhOP2}
\end{eqnarray}
Replacing all phonon operators by their averages, the Hamiltonian $H=H_{\rm e} + H_{\rm e-e} + H_{\rm e-ph} + H_{\rm ph} + H_{\rm ph-ph}$ becomes an effective electronic model, 
\begin{eqnarray}
\bar H &=& \sum_{\k,\alpha} g_{1\Q}(\k) 
(\delta_{\Q\alpha} c_{\k+\Q_\alpha \alpha}^\dagger f_\k^\pdagger
+ \delta_{\Q\alpha}^\ast f_\k^\dagger c_{\k+\Q_\alpha \alpha}^\pdagger )
\nonumber \\
&+&\frac{1}{2}\sum_\k \sum_{\alpha,\beta} [\bar A_\Q^f(\k) f_\k^\dagger f_\k^\pdagger
+ \bar A_\Q^c (\k) c_{\k \beta}^\dagger c_{\k \beta}^\pdagger]
[(\delta_{\Q\alpha}^\ast)^2 + \delta_{\Q\alpha}^2 ]
\nonumber \\
&+&\frac{1}{6}\sum_\k \sum_{\alpha,\beta} \{ \bar B_{1\Q}^{\alpha\beta}(\k) 
[(\delta_{\Q\beta}^\ast)^2 + \delta_{\Q\beta}^2] 
+ \bar B_{2\Q}^{\alpha\beta}(\k)|\delta_{\Q\beta}|^2\}
\nonumber \\
&\times& (\delta_{\Q\alpha} c_{\k+\Q_\alpha \alpha}^\dagger f_\k^\pdagger + 
\delta_{\Q\alpha}^\ast f_{\k}^\dagger c_{\k+\Q_\alpha \alpha}^\pdagger )
\nonumber \\
&+& \frac{1}{24} \sum_\k \sum_{\alpha,\gamma} \sum_{\beta\neq\alpha} \big\{ [ 
\widetilde C_{\alpha\beta}^f(\k) f_\k^\dagger f_\k^\pdagger 
+ \widetilde C_{\alpha\beta}^c (\k) c_{\k\gamma}^\dagger c_{\k\gamma}^\pdagger ] 
\nonumber \\
&\times& [(\delta_{\Q\alpha}^\ast \delta_{\Q\beta}^\ast)^2 
+ (\delta_{\Q\alpha}^\ast \delta_{\Q\beta})^2 ]
+ [(\widetilde C_{\alpha\beta}^f(\k))^\ast f_\k^\dagger f_\k^\pdagger 
\nonumber \\
&+& (\widetilde C_{\alpha\beta}^c (\k))^\ast c_{\k\gamma}^\dagger c_{\k\gamma}^\pdagger ] 
[(\delta_{\Q\alpha} \delta_{\Q\beta})^2+ (\delta_{\Q\alpha} \delta_{\Q\beta}^\ast)^2 ]\big\}
\nonumber \\
&+&\widehat DN\sum_\alpha \sum_{\beta>\alpha} [(\delta_{\Q\alpha}^\ast \delta_{\Q\beta}^\ast)^2 
+ (\delta_{\Q\alpha} \delta_{\Q\beta})^2 + (\delta_{\Q\alpha}^\ast \delta_{\Q\beta})^2
\nonumber \\
&+& (\delta_{\Q\alpha} \delta_{\Q\beta}^\ast)^2 ]
+\widetilde D N \sum_\alpha |\delta_{\Q\alpha}|^4 
+ \frac{\hbar \omega}{4} N\sum_\alpha |\delta_{\Q\alpha}|^2 
\nonumber \\
&+& \bar H_{\rm mEFKM}
,
\label{H_elec}
\end{eqnarray}
where $\alpha,\beta,\gamma=1,2,3$. Moreover, it is $\widehat D=2D(\Q_\alpha,\Q_\alpha,\Q_\beta,\Q_\beta)$, where $\beta\neq\alpha$, and $\widetilde D=D(\Q_\alpha,\Q_\alpha,\Q_\alpha,\Q_\alpha)$. The electron-phonon and phonon-phonon interaction constants are considered as real numbers except $\widetilde C_{\alpha\beta}^f(\k)$ and $\widetilde C_{\alpha\beta}^c (\k)$. The phases of these constants must be taken into account, otherwise chirality will not develop in our model. 
For the electron-phonon coupling constants we use the shorthand notation,
\begin{eqnarray}
g_{1\Q}(\k) &=& g_1(\k,\Q_\alpha,f,\alpha) ,
\\
\bar A_{1\Q}^f(\k) &=& g_2(\k,\Q_\alpha,\Q_\alpha,f,f) ,
\\
\bar A_{1\Q}^c(\k) &=& g_2(\k,\Q_\alpha,\Q_\alpha,1,1) ,
\\
\bar B_{1\Q}^{\alpha\beta}(\k) &=& 3 g_3(\k,\Q_\beta,\Q_\beta,\Q_\alpha,f,\alpha) ,
\\
\bar B_{2\Q}^{\alpha\beta}(\k) &=& 6 g_3(\k,\Q_\beta,-\Q_\beta,\Q_\alpha,f,\alpha) ,
\\
\widetilde C_{\alpha \beta}^f(\k) &=& 6 g_4(\k,\Q_\alpha,\Q_\alpha,\Q_\beta,\Q_\beta,f,f) ,
\\
\widetilde C_{\alpha \beta}^c (\k) &=& 6 g_4(\k,\Q_\alpha,\Q_\alpha,\Q_\beta,\Q_\beta,1,1) .
\end{eqnarray}

We assume for simplicity that the phonon-phonon interaction constants are the same for all combinations of $\alpha$ and $\beta$. In Eq.~\eqref{H_elec}, the term proportional to $\widetilde D$, coming from the expansion of the phonon-phonon interaction, guarantees that the free energy is bounded from below within our approximations.
Obviously, a finite lattice distortion  causes a hybridization between the valence and the conduction bands. As a consequence a gap in the electronic spectrum opens, just as in the course of exciton condensation [cf. Eq.~\eqref{H_dec}]. The corresponding local electron density is given by Eq.~\eqref{locDens}.

\begin{figure}[h]
\centering
\includegraphics[width=0.5\linewidth]{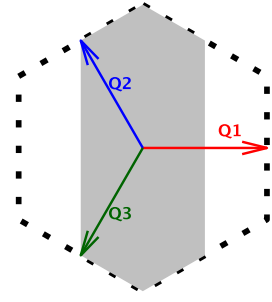}
\caption{(color online) BZ in the normal phase (black dotted hexagon) and (artificial)  BZ that would emerge if only the 
phonon $\Q_1$ softens (filled gray hexagon). Red, green, and blue arrows indicate the ordering vectors $\Q_1$, $\Q_2$, and $\Q_3$, respectively.}
\label{BZ_test}
\end{figure}

Of particular interest are  the phases $\theta_\alpha$. Owing to the terms on the r.h.s. of Eq.~\eqref{H_elec} proportional to $g_{1\Q}(\k)$ and $\bar B_{i\Q}^{\alpha\beta}(\k)$, these phases are coupled to the phases of $\delta_{\Q \alpha}$. Let us analyze the possible values of the phases of the static lattice distortion. We first note that every $\Q_\alpha$ is half a reciprocal lattice vector in the normal phase, i.e., $e^{2i\Q_\alpha {\bf R}_i}=1$, where ${\bf R}_i$ is a lattice vector of the undistorted lattice. Hence
\begin{eqnarray}
b_{\Q_\alpha}^\dagger &=& \frac{1}{\sqrt{N}} \sum_i b_i^\dagger e^{-i \Q_\alpha {\bf R}_i} 
= \frac{1}{\sqrt{N}} \sum_i b_i^\dagger e^{-i \Q_\alpha {\bf R}_i + 2i\Q_\alpha {\bf R}_i}\nonumber \\
&=& \frac{1}{\sqrt{N}} \sum_i b_i^\dagger e^{i \Q_\alpha {\bf R}_i} 
= b_{-\Q_\alpha}^\dagger \,.
\label{bq}
\end{eqnarray}
That is, $b_{\Q\alpha}^\dagger$ and $b_{-\Q_\alpha}^\dagger$ create the same phonon. This implies $\langle b_{\Q_\alpha}^\dagger \rangle = \langle b_{-\Q_\alpha}^\dagger \rangle = \langle b_{\Q_\alpha}^\pdagger \rangle$. Consequently, $\langle b_{\Q_\alpha}\rangle$ and $\delta_{\Q\alpha}$ become real  numbers.
However, since a triple CDW is not a simple superposition of three single CDWs,  the situation is more subtle. Here, the change of the periodicity of the lattice caused by one CDW component affects the formation of the other two components. To elucidate  this in more detail let us assume that phonon $1$ softens at $T_{c}$, while phonon $2$ and phonon $3$ soften at $T_{c}-\delta T$. As a result of the transition $1$ at $T_c$ the periodicity of the crystal changes and consequently the BZ changes too (cf. Fig.~\ref{BZ_test}). 
The vectors $\Q_2$ and $\Q_3$ are no longer half-reciprocal lattice vectors, and Eq.~\eqref{bq} does not apply. Hence, at $T_c-\delta T$, $\langle b_{\Q 2}\rangle$ and $\langle b_{\Q 3}\rangle$ are complex numbers with phases that have to be determined by minimizing the free energy. For 1$T$-TiSe$_2$, $\delta T=0$, but nevertheless the above discussion remains valid. That is the freedom to fix the phases of the lattice distortions in an appropriate way results from the fact that one triple-CDW component must develop in a lattice structure which is already distorted by the other two triple-CDW components.

\subsection{Ground-state energy}
\label{sec:GS_energy}
Based on the model~\eqref{H_elec} we analyze the chiral CDW formation at zero temperature. Taking into account the symmetry of the conduction bands and the equality of the interaction constants, we have $|\delta_{\Q1}|=|\delta_{\Q2}|=|\delta_{\Q3}|=|\delta_{\Q}|$ and $|\Delta_{\Q1}|=|\Delta_{\Q2}|=|\Delta_{\Q3}|=|\Delta_{\Q}|$. 
Therewith the ground-state energy per site follows as 
\begin{align}
\frac{\bar E}{N}&= \frac{2}{N}\sum_{\k,\alpha} g_{1\Q}(\k) |\delta_{\Q}| |\langle c_{\k+\Q_\alpha \alpha}^\dagger f_\k^\pdagger \rangle| \cos\left(\phi_\alpha-\theta_\alpha \right)
\nonumber \\
&+\frac{1}{N}\sum_\k \sum_{\alpha,\beta} |\delta_{\Q}|^2 
[\bar A_\Q^f(\k) \langle f_\k^\dagger f_\k^\pdagger \rangle 
+ \bar A_\Q^c(\k) \langle c_{\k\beta}^\dagger c_{\k\beta}^\pdagger \rangle ] 
\nonumber \\ 
&\times\cos(2\phi_\alpha)
+\frac{1}{3N} \sum_\k \sum_{\alpha,\beta} |\delta_{\Q}|^3 
|\langle c_{\k+\Q_\alpha \alpha}^\dagger f_\k^\pdagger \rangle |
\nonumber \\
&\times [2\bar B_{1\Q}^{\alpha\beta}(\k) \cos(2\phi_\beta) + \bar B_{2\Q}^{\alpha \beta}(\k)]
\cos(\phi_\alpha-\theta_\alpha)
\nonumber \\
&+\frac{1}{12N}\sum_\k \sum_{\alpha,\gamma} \sum_{\beta\neq\alpha}
[\bar C_{\alpha\beta}^f(\k) \langle f_\k^\dagger f_\k^\pdagger \rangle 
+ \bar C_{\alpha\beta}^c(\k)\langle c_{\k\gamma}^\dagger c_{\k\gamma}^\pdagger \rangle]
\nonumber \\
&\times  |\delta_{\Q}|^4 \big[ \cos(2(\phi_\alpha-\phi_\beta)+\phi_C)
+ \cos(2(\phi_\alpha+\phi_\beta)+\phi_C)\big]
\nonumber \\
&+4\widehat D \sum_\alpha \sum_{\beta>\alpha} |\delta_{\Q}|^4 
\cos(2\phi_\alpha) \cos(2\phi_\beta) 
+ 3\widetilde D |\delta_{\Q}|^4
\nonumber \\
&+\frac{3}{4}\hbar \omega |\delta_{\Q}|^2
+\frac{\bar E_{\rm mEFKM}}{N} \,,
\label{E}
\end{align}
where $\widetilde C_{\alpha\beta}^{f(c)} = \bar C_{\alpha\beta}^{f(c)} e^{-i\phi_C}$.
We note that each phase $\theta_\alpha$ is exclusively coupled to $\phi_\alpha$.  
If $\sum_{\k} \big[ g_{1\Q}(\k)+\sum_{\beta} (2\bar B_{1\Q}^{\alpha\beta}(\k) \cos(2\phi_\beta) +\bar B_{2\Q}^{\alpha\beta}(\k) )|\delta_{\Q\beta}|^2 \big] 
|\langle c_{\k+\Q_\alpha \alpha}^\dagger f_\k^\pdagger \rangle | >0 $,
the choice
\begin{equation}
\theta_\alpha = \phi_\alpha+(2s+1)\pi ,
\label{theta_phi}
\end{equation}
minimizes the energy, where $s=0,1,2,...$ Otherwise $\theta_\alpha$ are locked to $\phi_\alpha+2s\pi$. 
Thus, the relationship between $\phi_1$, $\phi_2$, and $\phi_3$ is crucial.

The Hamiltonian~\eqref{H_elec} offers a complex model with many (unknown) parameters. To proceed we assume the electron-phonon interaction constants as independent of the momentum $\k$. Moreover, we assume  that $\bar A_\Q^f$, $\bar A_\Q^c$, $\bar B_{1\Q}^{\alpha\beta}$, $\bar B_{2\Q}^{\alpha\beta}$, $\bar C_{\alpha\beta}^f$, $\bar C_{\alpha\beta}^c$, $\widehat D$, and $\widetilde D$ are much smaller than $U_{fc}$, $U_{cc}$, and $g_{1\Q}$. The magnitude of the EI order parameter and the static lattice distortion are then primarily determined by the latter interaction constants and the constraint for Eq.~\eqref{theta_phi} simply reduces to $g_{1\Q}>0$. 
Taking only $U_{fc}$, $U_{cc}$, and $g_{1\Q}$ into account and minimizing the free energy with respect to the EI order parameter yields $\frac{\partial F}{\partial |\Delta_\Q|}=\frac{\partial F}{\partial |\widetilde\Delta_\Q|}+\frac{6}{U_{fc}} |\Delta_\Q|=0$, while the minimization with respect to the static lattice distortion yields $\frac{\partial F}{\partial |\delta_\Q|}=g_{1\Q} \frac{\partial F}{\partial |\widetilde\Delta_\Q|}+\frac{6}{4}\hbar\omega |\delta_\Q|=0$, where the gap parameter is given by
\begin{equation}
\widetilde \Delta_{\Q\alpha} = g_{1\Q} \delta_{\Q\alpha}-\Delta_{\Q\alpha} .
\label{gap_parameter}
\end{equation}
The relation~\eqref{theta_phi} maximizes the modulus of the gap parameter (supposing $g_{1\Q}>0$).
From the energy minimization with respect to $|\Delta_\Q|$ and $|\delta_\Q|$ 
\begin{equation}
|\Delta_\Q|=\frac{U_{fc}}{4}\frac{\hbar\omega}{g_{1\Q}} |\delta_\Q| .
\label{Delta_delta}
\end{equation}
With Eqs.~\eqref{theta_phi} and \eqref{Delta_delta} we can express the ground-state energy per site as
\begin{equation}
\frac{\bar E}{N}=\frac{1}{N} \bar E_\delta(|\delta_\Q|^2) 
+ \frac{1}{N} \bar E_\phi(|\delta_\Q|^2, \phi_1, \phi_2, \phi_3) ,
\label{E_exp}
\end{equation}
where
\begin{align}
\frac{1}{N}\bar E_\delta & (|\delta_\Q|^2) =
\frac{1}{N}\sum_{\k,\nu} E_{\k\nu} \langle n_{\k}^{\nu}\rangle - U_{fc} n_f (1-n_f) 
\nonumber \\
&- \frac{U_{cc}}{3} (1-n_f)^2  + \frac{3}{4}\hbar\omega |\delta_\Q|^2 + \frac{3}{16} U_{fc} \bigg(\frac{\hbar\omega}{g_{1\Q}}\bigg)^2 |\delta_\Q|^2 
\nonumber \\
&+ \bigg(3\widetilde D - \frac{1}{12} \widehat B_{2\Q} 
\frac{\hbar\omega}{g_{1\Q}} \bigg) |\delta_\Q|^4 ,
\label{E_delta}
\\
\frac{1}{N}\bar E_\phi & (|\delta_\Q|^2,\phi_1,\phi_2,\phi_3) =
\big[\bar A_\Q^f n_f + \bar A_\Q^c (1-n_f)\big] |\delta_\Q|^2 
\nonumber \\
&\times  \sum_\alpha \cos(2\phi_\alpha) - \frac{1}{6} \frac{\hbar\omega}{g_{1\Q}} 
|\delta_\Q|^4 \sum_\alpha \widehat B_{1\Q}^\alpha \cos(2\phi_\alpha) 
\nonumber \\
&+ \frac{1}{12} \sum_\alpha \sum_{\beta\neq\alpha} 
\big[ \bar C_{\alpha\beta}^f n_f + \bar C_{\alpha\beta}^c (1-n_f)\big] |\delta_\Q|^4
\nonumber \\
&\times \big[ \cos(2(\phi_\alpha-\phi_\beta)+\phi_C) 
+ \cos(2(\phi_\alpha+\phi_\beta)+\phi_C)\big]
\nonumber \\
&+4\widehat D |\delta_\Q|^4 \sum_\alpha \sum_{\beta>\alpha} \cos(2\phi_\alpha)\cos(2\phi_\beta) .
\label{E_phi}
\end{align}
Here, $\widehat B_{2\Q}=\sum_{\alpha,\beta} \bar B_{2\Q}^{\alpha\beta}$ and
$\widehat B_{1\Q}^\alpha = \sum_\beta \bar B_{1\Q}^{\beta\alpha}$. The quasiparticle energies $E_{\k\nu}$ ($\nu=A,B,C,D$) are obtained by the diagonalization of the Hamilton matrix
\begin{equation}
[H] = \begin{pmatrix} \beps_{\k f} & \tilde\Delta_{\Q 1}^\ast & \tilde\Delta_{\Q 2}^\ast & \tilde\Delta_{\Q 3}^\ast \\ 
\tilde\Delta_{\Q 1} & \beps_{\k+\Q_1 1} & 0 & 0 \\
\tilde\Delta_{\Q 2} & 0 & \beps_{\k+\Q_2 2} & 0 \\
\tilde\Delta_{\Q 3} & 0 & 0 & \beps_{\k+\Q_3 3} \end{pmatrix} .
\label{matrixH}
\end{equation}
Since only $|\widetilde\Delta_\Q|^2$ enters $E_{\k\nu}$ we may replace $\widetilde\Delta_{\Q\alpha}$ by
\begin{equation}
|\widetilde\Delta_\Q| = \bigg( g_{1\Q}+\frac{U_{fc}}{4} \frac{\hbar\omega}{g_{1\Q}} \bigg)
|\delta_\Q|
\end{equation}
in Eq.~\eqref{matrixH}. The choice 
\begin{eqnarray}
\widetilde D &\geq& 4\widehat D +\frac{1}{3}\sum_{\alpha}\sum_{\beta>\alpha} \big(\bar C_{\alpha\beta}^f + \bar C_{\alpha\beta}^c \big) +\frac{1}{6}\sum_\alpha \widehat B_{1\Q}^\alpha \frac{\hbar\omega}{g_{1\Q}} 
\nonumber \\
&&+ \frac{1}{12} B_{2\Q} \frac{\hbar\omega}{g_{1\Q}}
\label{tildeD}
\end{eqnarray}
guarantees the lower boundary of the energy. In the numerical calculation we use the equality in Eq.~\eqref{tildeD}.

Only the electron-phonon interaction and the phonon-phonon interaction enter the phase-dependent part of the ground-state energy $\bar E_\phi$.
It is the quartic order expansion term of the electron-phonon interaction and the phonon-phonon interaction (also in quartic order of the lattice distortion) that relate the phases $\phi_1$, $\phi_2$, and $\phi_3$ to each other and favor a phase difference. Without them chirality can not be stabilized.
Note that the $2\times2\times2$-commensurability of the CDW is an important prerequisite for $\bar E_\phi\neq 0$. This rules out incommensurate CDWs exhibiting a chiral property.

The chiral CDW can be indicated by
\begin{equation}
d_\phi = |\delta_\Q| |(\phi_1-\phi_2)(\phi_1-\phi_3)(\phi_2-\phi_3)| .
\label{d_phi}
\end{equation}
$d_\phi$ is finite only if the CDW is realized and $\phi_1\neq\phi_2\neq\phi_3$; i.e., it fulfills a prerequisite for an order parameter of the chiral CDW.

\subsection{Phase boundary of the CDW}
\label{sec:Ph_boundary}
In contrast to the ground-state energy~\eqref{E_exp} the constraint for the CDW phase boundary can be obtained from Eq.~\eqref{H_elec} without approximations. The derivative of the free energy with respect to the static lattice distortion is needed in the limit $|\delta_\Q|\rightarrow 0$. To this end, we use the ansatz 
\begin{eqnarray}
E_{\k A} &=& \beps_{\k f} + \sum_\alpha |\delta_\Q|^2 \bar A_\Q^f \cos(2\phi_\alpha)
+|\widetilde\Delta_\Q|^2 d_A
\nonumber \\
&+&\frac{1}{12}\sum_\alpha \sum_{\beta\neq\alpha} |\delta_\Q|^4 
\bar C_{\alpha\beta}^f \Big[ \cos(2(\phi_\alpha-\phi_\beta)+\phi_C) 
\nonumber \\
&+& \cos(2(\phi_\alpha+\phi_\beta)+\phi_C) \Big] ,
\label{exp_Eka}
\end{eqnarray}
\begin{eqnarray}
E_{\k B} &=& \beps_{\k+\Q_1 1} + \sum_\alpha |\delta_\Q|^2 \bar A_\Q^c \cos(2\phi_\alpha)
+|\widetilde\Delta_\Q|^2 d_B
\nonumber \\
&+&\frac{1}{12}\sum_\alpha \sum_{\beta\neq\alpha} |\delta_\Q|^4
\bar C_{\alpha\beta}^c \Big[ \cos(2(\phi_\alpha-\phi_\beta)+\phi_C) 
\nonumber \\
&+& \cos(2(\phi_\alpha+\phi_\beta)+\phi_C) \Big] ,
\label{exp_Ekb}
\end{eqnarray}
\begin{eqnarray}
E_{\k C} &=& \beps_{\k+\Q_2 2} + \sum_\alpha |\delta_\Q|^2 \bar A_\Q^c \cos(2\phi_\alpha)
+|\widetilde\Delta_\Q|^2 d_C
\nonumber \\
&+&\frac{1}{12}\sum_\alpha \sum_{\beta\neq\alpha} |\delta_\Q|^4
\bar C_{\alpha\beta}^c \Big[ \cos(2(\phi_\alpha-\phi_\beta)+\phi_C) 
\nonumber \\
&+& \cos(2(\phi_\alpha+\phi_\beta)+\phi_C) \Big] , 
\label{exp_Ekc}
\end{eqnarray}
\begin{eqnarray}
E_{\k D} &=& \beps_{\k+\Q_3 3} + \sum_\alpha |\delta_\Q|^2 \bar A_\Q^c \cos(2\phi_\alpha)
+|\widetilde\Delta_\Q|^2 d_D
\nonumber \\
&+&\frac{1}{12}\sum_\alpha \sum_{\beta\neq\alpha} |\delta_\Q|^4
\bar C_{\alpha\beta}^c \Big[ \cos(2(\phi_\alpha-\phi_\beta)+\phi_C) 
\nonumber \\
&+& \cos(2(\phi_\alpha+\phi_\beta)+\phi_C) \Big] .
\label{exp_Ekd}
\end{eqnarray}
The unknown parameters $d_\nu$, $\nu=A,B,C,D$ can be calculated from the characteristic polynomial of the Hamilton matrix. With the ansatz Eqs.~\eqref{exp_Eka}-\eqref{exp_Ekd} and Eq.~\eqref{Delta_delta}, which also holds up to linear order in the static lattice distortion, the free energy can be minimized analytically, which gives the exact result in the limit $|\delta_\Q|\rightarrow 0$. Considering this limit the constraint for the CDW phase boundary is obtained as
\begin{eqnarray}
0 &=&\frac{3}{4}\hbar\omega g_{1\Q}^2 + \frac{3}{16}U_{fc}(\hbar\omega)^2 - 3 g_{1\Q}^2 \big[\bar A_\Q^f n_f + \bar A_\Q^c (1-n_f)\big]
\nonumber \\
&&+ \bigg( g_{1\Q}^2+\frac{1}{4}U_{fc}\hbar\omega \bigg)^2 \frac{1}{N}\sum_\k \bigg( \frac{\bar m_{E_{\k A}}}{\bar n_{E_{\k A}}} \langle n_{\k}^f \rangle 
\nonumber \\
&&+ \frac{\langle n_{\k+\Q_1}^1\rangle}{\beps_{\k+\Q_1 1}-\beps_{\k f}} + \frac{\langle n_{\k+\Q_2}^2 \rangle}{\beps_{\k+\Q_2 2}-\beps_{\k f}} + \frac{\langle n_{\k+\Q_3}^3 \rangle}{\beps_{\k+\Q_3 3}-\beps_{\k f}} \bigg) ,
\nonumber \\
\label{phase_boundary}
\end{eqnarray}
where
\begin{eqnarray}
\bar m_{E_{\k A}} &=&\beps_{\k+\Q_1 1} \beps_{\k+\Q_2 2} + \beps_{\k+\Q_1 1} \beps_{\k+\Q_3 3}
+\beps_{\k+\Q_2 2} \beps_{\k+\Q_3 3} 
\nonumber \\
&&+3 \beps_{\k f}^2 - 2\beps_{\k f} (\beps_{\k+\Q_1 1}
+\beps_{\k+\Q_2 2} + \beps_{\k+\Q_3 3}) ,
\\
\bar n_{E_{\k A}} &=& \beps_{\k f}^3 - \beps_{\k f}^2(\beps_{\k+\Q_1 1} + \beps_{\k+\Q_2 2}
+\beps_{\k+\Q_3 3}) 
\nonumber \\
&&+\beps_{\k f}(\beps_{\k+\Q_1 1}\beps_{\k+\Q_2 2} + \beps_{\k+\Q_1 1}
\beps_{\k+\Q_3 3} 
\nonumber \\
&&+ \beps_{\k+\Q_2 2} \beps_{\k+\Q_3 3} )
-\beps_{\k+\Q_1 1} \beps_{\k+\Q_2 2} \beps_{\k+\Q_3 3} .
\label{constraint_PB}
\end{eqnarray}

\subsection{Characterization of the CDW state in 1$T$-TiSe$_2$ }
\label{sec:PhTransitionChar}
\begin{figure}[h]
\centering
\subfigure[]{\includegraphics[width=0.49\linewidth]{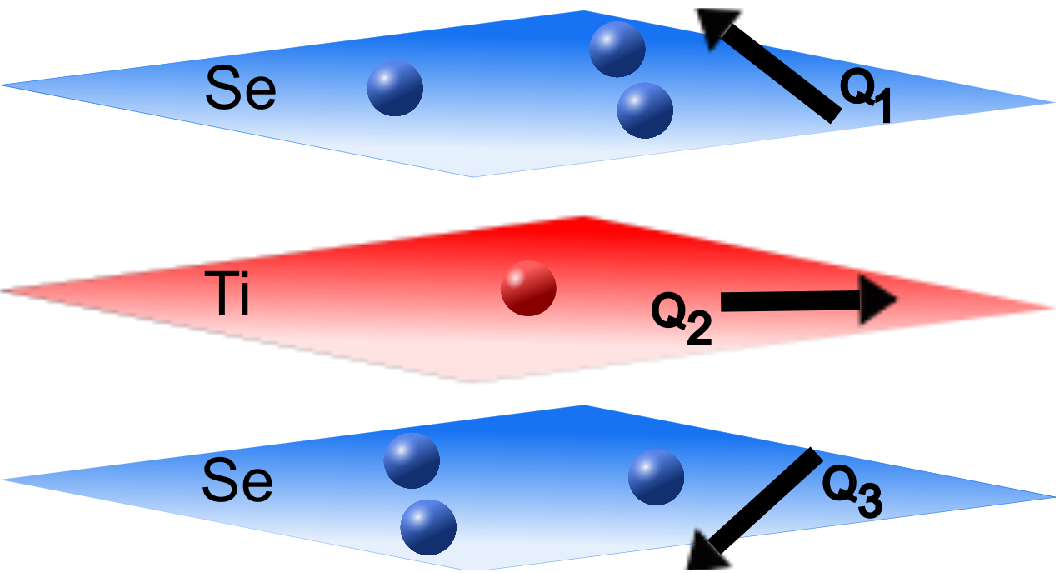}}
\subfigure[]{\includegraphics[width=0.49\linewidth]{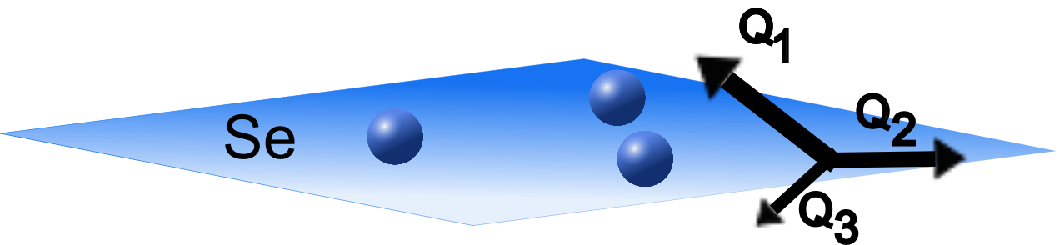}}
\caption{(color online) (a) For a chiral ordering the maximum  lattice distortion due to the  phonon $\Q_\alpha$ may be located in different ionic layers. (b) The Se ions in the upper layer are differently affected by the phonons having  momentum $\Q_1$, $\Q_2$, or $\Q_3$. 
For further discussion see text.}
\label{fig:PLD_sketch}
\end{figure}

Experiments identify  a  close connection between the appearance of the CDW state and the periodic lattice displacement  in 1$T$-TiSe$_2$.\cite{We11} 
The displacement of the ion $m$ in the unit cell $n$ is
\begin{equation}
\widetilde u(n,m) = \sum_\alpha \frac{\hbar}{\sqrt{2M_m \omega}} |\delta_\Q| {\bf \epsilon}(m,\Q_\alpha) \cos(\Q_\alpha R_n - \phi_\alpha),
\label{latt_deform}
\end{equation}
where ${\bf \epsilon}(m,\Q_\alpha)$ is the polarization vector and $M_m$ is the mass of the ion $m$.
Clearly each CDW component $\alpha$ produces a 3D lattice distortion. If $\phi_1\neq\phi_2\neq\phi_3$, the lattice will be differently affected by the phonons $\Q_1$, $\Q_2$, and $\Q_3$. 
Of course the lattice deformation by the phonon mode $\Q_\alpha$ is position-dependent; in this way  a complicated distortion pattern of the ions can occur. An instructive picture can be achieved, however, if one neglects the position-dependence in the $xy$-plane. In this simplified situation, depending on the $z$-component as a function of the position, the magnitude of the lattice displacement differs along $\Q_1$, $\Q_2$, and $\Q_3$. As a result the different ionic layers of 1$T$-TiSe$_2$ are dominated by different phonon modes.\cite{We11} The situation where  the lower Se-ion layer is largely affected by the phonon mode $\Q_3$, the Ti-ion layer by phonons with momentum $\Q_2$, and the upper Se-ion layer by the  $\Q_1$ phonon mode, is illustrated schematically   in Fig.~\ref{fig:PLD_sketch}(a). 
Let us consider the upper plane of Se ions, which is analyzed in scanning tunneling microscopy  experiments.\cite{ILSKITOT10} There, a relative difference of the phases $\phi_{\alpha}$ leads, e.g.,  to a stronger displacement of the ions in the direction of $\Q_1$ than in the direction of $\Q_2$ and $\Q_3$ [see Fig.~\ref{fig:PLD_sketch}(b)].  Then the CDW transition can be viewed as the formation of ``virtual layers'' with ordering vectors assigned to a helical structure.\cite{ILSKITOT10}  This distortion scenario equates with a fixed phase difference. Thereby the only crucial parameters are $\phi_1$, $\phi_2$, and $\phi_3$; the finite $z$ component of the ordering vectors is not a required prerequisite for the chiral CDW. 
Although the different orbital character of the CDW components do not directly influence the charge modulation, the phase difference leads to a different transfer of spectral weight along $\Q_1$, $\Q_2$, and $\Q_3$ and the formation of a chiral CDW necessarily generates an orbital-ordered state.~\cite{We12}

Equation~\eqref{E}  specifies values for the phases $\theta_\alpha$ and $\phi_\alpha$. Which particular phase takes one of these values remains open. For instance, the simultaneous transformations $\theta_2\rightarrow\theta_3$ and $\phi_2\rightarrow\phi_3$ do not change the energy, but convert a clockwise chiral CDW in an anticlockwise one. The degeneracy of these two CDW states is in accord with the experimental findings for 1$T$-TiSe$_2$.\cite{ILSKITOT10}

As it is apparent in Eq.~\eqref{Delta_delta}, for finite $U_{fc}$ and $g_{1\Q}$ the EI order parameter $|\Delta_\Q|>0$ if and only if $|\delta_\Q|>0$. This can also be argued on physical grounds. Let us first consider the case of vanishing electron-phonon coupling. In the EI phase ($|\Delta_\Q|> 0$) the system realizes a CDW. When $g_{1\Q}$ becomes finite in addition, the lattice adjusts commensurate with the  electron density modulation. Hence, in this case, any finite $g_{1\Q}$ immediately results in $|\delta_\Q|>0$.   
On the other hand, at vanishing Coulomb interaction but sufficiently large $g_{1\Q}>g_{1\Q,c}$, a lattice instability develops leading to a finite $\sum_\k \langle c_{\k+\Q_\alpha \alpha}^\dagger f_\k^\pdagger \rangle$. This hybridization parameter enters the explicit equation for the EI order parameter, see Eq.~\eqref{EIOP_def}. $|\Delta_\Q|>0$ then follows from any finite Coulomb interaction. Our approach therefore does not  discriminate between an excitonic and phonon-driven instability if both electron-electron and electron-phonon interactions are at play.

\section{Numerical results}
\label{sec:NumericalResults}

\subsection{Model assumptions}
\label{sec:mod_params}
In view of the quasi-2D crystallographic and electronic structure of 1$T$-TiSe$_2$,  and in order to simplify the numerics,
we restrict  the  following analysis to a strictly 2D setting. Moreover, being close to the Fermi energy, we will approximate     
the bands parabolically:\cite{MCCBSDGBA09} 
\begin{eqnarray}
\varepsilon_{\k f} &=& -t_f \left( k_x^2+k_y^2\right),
\label{eqn_VB}\\
\varepsilon_{\k 1} &=& t_{c}^x (k_x-Q_{1x})^2 + t_{c}^y (k_y-Q_{1y})^2 +E_c , 
\label{eqn:CB1}
\end{eqnarray}
with hopping amplitudes $t_f$, $t_c^x$, and $t_c^y$.  The other two  conduction bands 
$\varepsilon_{\k 2}$ and $\varepsilon_{\k 3}$ have analogous dispersions, but 
the momenta  are rotated by $2\pi/3$ and $4\pi/3$, respectively. 
All three conduction bands share the same minimum $E_c$,  see Fig.~\ref{fig:bands}.  

From the band dispersion provided by Monney {\it et al.} in Ref.~\onlinecite{MCCBSDGBA09} we derive $t_f=1.3$ eV, which will be taken as the unit of energy hereafter,  and $t_c^x=0.042$ and $t_c^y=0.105$. The bare phonon frequency is estimated  as $\hbar \omega=0.013$, in accordance with the value given by Weber {\it et al.} in Ref.~\onlinecite{WRCOKHHBA11}. Furthermore, we set $E_c=-3.30$ and $U_{cc}=U_{fc}+1.0$. Note that $E_c$ is the minimum of the bare conduction band. The effective band overlap will be significantly smaller due to the Coulomb interaction induced  Hartree shift. 
If it is not explicitly noted we take $\widehat B_{1\Q}^\alpha=0.5\times 10^{-4}$, $\widehat B_{2\Q}=10^{-4}$, $\bar C_{\alpha\beta}^f=\bar C_{\alpha\beta}^c=8.5\times 10^{-4}$, $\widehat D=10^{-5}$, and $\phi_C=3\pi/10$.

The self-consistency loop, comprising the determination of the total and partial
particle densities and the chemical potential, is solved iteratively until the  relative error of each physical quantity is less than 10$^{-6}$.  
The numerical integrations were performed using the  Cubpack package.\cite{CH03}

\subsection{Formation scenario of the chiral CDW}
\label{sec:chiral}
\begin{figure}[h]
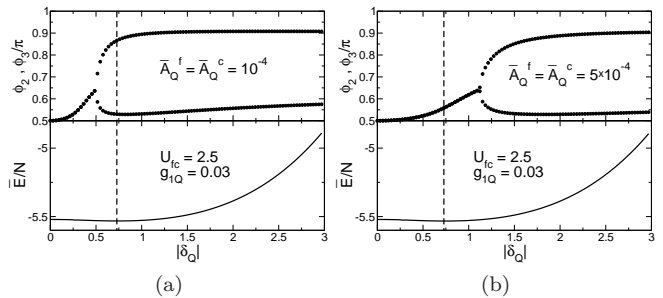

\centering
\subfigure[]{\includegraphics[width=0.49\linewidth]{fig6a_parameters1.eps}}
\subfigure[]{\includegraphics[width=0.49\linewidth]{fig6b_parameters3.eps}}
\caption{(Upper panels) Phases and (lower panels) ground-state energy as a function of the static lattice distortion. The dashed lines designate the physical value of the static lattice distortion. We set $\phi_1=\pi/2$ and the interband Coulomb interaction is $U_{fc}=2.5$. For the electron-phonon interaction we take $g_{1\Q}=0.03$, (a) $\bar A_\Q^f=\bar A_\Q^c=10^{-4}$, and (b) $\bar A_\Q^f=\bar A_\Q^c=5\times 10^{-4}$.}
\label{fig:OP_expa}
\end{figure}

We start with the analysis of the ground-state energy ($T=0$), where we treat the static lattice distortion as a variational parameter. Without loss of generality we choose $\phi_1=\pi/2$. The other phases $\phi_2$ and $\phi_3$ are determined by minimizing $\bar E_\phi /N$ using a simplex method. The results for $U_{fc}=2.5$ and $g_{1\Q}=0.03$ are shown in Fig.~\ref{fig:OP_expa}.

Since we assumed the nonlinear electron-phonon and the phonon-phonon interaction constants are much smaller than $U_{fc}$, $U_{cc}$, and $g_{1\Q}$, the energy  $\bar E/N \approx \bar E_\delta /N$ and the (physical) static lattice distortion, given by the dashed lines in Figs.~\ref{fig:OP_expa}, is primarily determined by the Coulomb interaction and $g_{1\Q}$.

We find a complex formation scenario for the chiral property. For $|\delta_Q|\rightarrow 0$ all phases are equal, i.e., $\phi_1=\phi_2=\phi_3=\pi/2$ and the CDW is nonchiral. With growing static lattice distortion $\phi_2=\phi_3\neq \phi_1$. Compared with the normal phase and the limit $|\delta_Q|\rightarrow 0$ the mirror symmetry is reduced in this state. However, there exists a mirror symmetry along $\Q_1$ (cf. Fig.~\ref{fig:mirror_sym}), and the CDW is still nonchiral. If the static lattice distortion exceeds a threshold, chirality sets in and $\phi_1\neq\phi_2\neq\phi_3$.

\begin{figure}[h]
\centering
\includegraphics[width=\linewidth]{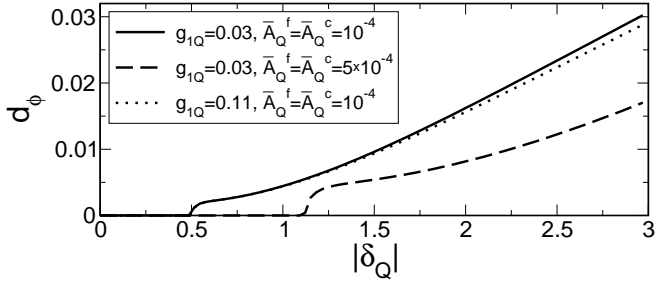}
\caption{Chiral-CDW characteristic quantity $d_\phi$ as a function of the static lattice distortion. The solid line represents the result for $g_{1\Q}=0.03$ and $\bar A_\Q^f=\bar A_\Q^c=10^{-4}$, the long-dashed line is the result for $g_{1\Q}=0.03$ and $\bar A_\Q^f=\bar A_\Q^c=5\times 10^{-4}$, and the dotted line shows the result for $g_{1\Q}=0.11$ and $\bar A_\Q^f=\bar A_\Q^c=10^{-4}$.}
\label{fig:d_phi}
\end{figure}

With increasing $\bar A_\Q^{f(c)}$ the threshold for the static lattice distortion that separates the chiral and the nonchiral CDW grows, see Fig.~\ref{fig:d_phi}. The electron-phonon interaction constant $g_{1\Q}$ barely influences the chiral property of the CDW.

The scenario shown in Fig.~\ref{fig:OP_expa} suggests that coming from the uniform, high temperature phase and lowering $T$ there is first a  transition to the nonchiral CDW at $T_{\rm nonchiral\, CDW}$. Chirality is formed at $T_{\rm chiral\, CDW}< T_{\rm nonchiral\, CDW}$. 
This sequence of transitions agrees with the result from the Landau-Ginzburg approach~\cite{We11,We12} and is supported by very recent x-ray diffraction and electrical transport measurements.\cite{CROLGKRW12} The difference between $T_{\rm nonchiral\, CDW}$ and $T_{\rm chiral\,CDW}$ is estimated experimentally to be less than 10~K.  
Moreover, the suggested transition scenario does not contradict experiment, where 1$T$-TiSe$_2$ is gradually doped with Cu until the CDW is suppressed in favor of a superconducting phase.\cite{ICZGMK12} Here, chirality is present until the breakdown of the CDW. Since the transition from the CDW to the superconducting phase is affirmed as a first order transition,\cite{MZDBOKROC06} $|\delta_\Q|$ does not have to be small at the phase boundary and chirality may exist.

\begin{figure}[h]
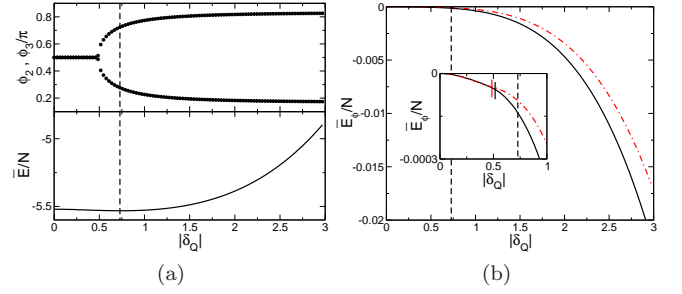

\centering
\subfigure[]
{\includegraphics[height=0.385\linewidth]{fig8a_parameters1_vW.eps}}
\subfigure[]
{\includegraphics[height=0.385\linewidth]{fig8b_weVSvW_parameters1.eps}}
\caption{(color online) (a) Phases and ground-state energy as a function of the static lattice distortion, where the functional dependency on the phases $\phi_\alpha$ is assumed as in Refs.~\onlinecite{We11,We12}. (b) Comparison of the phase-dependent part of the ground-state energy Eq.~\eqref{E_phi} (black solid line) and the counterpart for the phase dependency as suggested by van Wezel~\cite{We11,We12} (red dot-dashed line). The small vertical lines in the inset indicate the critical $|\delta_\Q|$ for the onset of chirality in the respective approximation scheme. In both figures the dashed line designates the physical value of the static lattice distortion. The model parameters are $U_{fc}=2.5$, $g_{1\Q}=0.03$, $\bar A_\Q^f=\bar A_\Q^c=10^{-4}$.}
\label{fig:OP_expa_vW}
\end{figure}

To combine our approach with the Landau-Ginzburg treatment we set $\widehat B_{1\Q}^\alpha =0$, $\widehat D=0$, neglect the terms $\cos(2(\phi_\alpha+\phi_\beta)+\phi_C)$, and set $\phi_C=0$.
Our model then reproduces the functional dependency of the free energy functional in Refs.~\onlinecite{We11,We12}. The Landau-Ginzburg parameters can then be expressed as 
\begin{align}
& \frac{3}{2} a_0 = -\frac{3}{4} \hbar\omega 
- \frac{3}{16}U_{fc}\bigg(\frac{\hbar\omega}{g_{1\Q}}\bigg)^2 , \label{LG_param1}
\\
& \frac{1}{2}a_1(1-\gamma) = \bar A_\Q^f n_f + \bar A_\Q^c (1-n_f) , \label{LG_param2}
\\
& \frac{3}{8}(15c_0+8d_0) = 3\widetilde D - \widehat B_{2\Q}\frac{\hbar\omega}{g_{1\Q}} ,\label{LG_param3} 
\\
& \frac{3}{4} c_2 = \bar C_{\alpha\beta}^f n_f + \bar C_{\alpha\beta}^c (1-n_f) . \label{LG_param4}
\end{align}
Figure~\ref{fig:OP_expa_vW}(a) shows an example for this scheme.
Note that the phases $\phi_\alpha$ are periodic with $\pi$ and Fig.~\ref{fig:OP_expa_vW}(a) shows that $\phi_2=-\phi_3$, which was obtained analytically in Refs.~\onlinecite{We11,We12}. Most notably, if the $\cos(2(\phi_\alpha+\phi_\beta)+\phi_C)$ contribution and the phase $\phi_C$ are neglected the ``intermediate'' state where $\phi_2=\phi_3\neq\phi_1$ is missing. The chiral CDW emerges directly from the nonchiral CDW, where $\phi_1=\phi_2=\phi_3=\pi/2$. The comparison of the phase-dependent part $\bar E_\phi /N$ shows that the approximation provided by Eq.~\eqref{E_phi} exhibits the lower energy. The onset of the chiral CDW differs only slightly between the two approximation schemes.

\subsection{Phase diagram of the mEFKM}
\label{sec:mEFKM}
\begin{figure}[h]
\centering
\includegraphics[width=\linewidth]{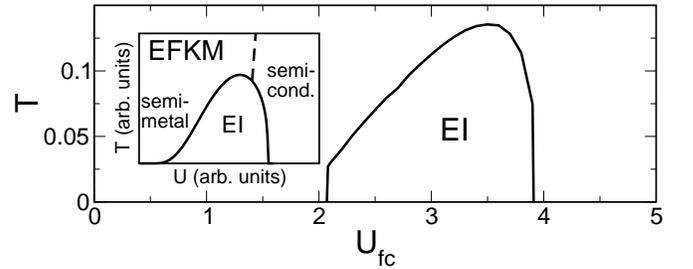}
\caption{Phase diagram of the mEFKM. The inset displays  the schematic phase diagram of the simplified two-band EFKM on a square lattice according to Ref.~\onlinecite{ZIBF12}.}
\label{fig:noPhonons}
\end{figure}

To set the stage for the analysis of the interplay of Coulomb and electron-phonon interaction effects we first discuss the phase diagram of the pure mEFKM, cf.~Fig.~\ref{fig:noPhonons}. 
Here, since $g_{1\Q}=0$ (and as a result $\delta_{\Q \alpha}=0$), the EI low-temperature phase typifies a normal CDW.  
As for the EFKM on a square lattice (see inset), at $T=0$ we  find a finite critical Coulomb strength above which the EI phase
does not exist. This is because the large band splitting caused by the Hartree term of the Coulomb interaction prevents $c$-$f$ 
electron coherence.\cite{ZIBF12}       
In contrast  to the EFKM, in our four-band model we also find a critical lower Coulomb strength for the EI phase. This can be understood as follows:  since the valence band is isotropic while the conduction-band dispersions are strongly anisotropic, particles close to the Fermi level do not find a large number of partners with appropriate momentum for electron-hole pairing. Thus, for 
$U_{fc}$ smaller than a critical Coulomb attraction, the amount of energy to create a macroscopic number of excitons is larger than the energy gain from the condensation transition into the EI state. Therefore the system remains in the semimetallic  phase.\cite{Zi67} 
The rather abrupt increase of the critical temperature at the lower critical Coulomb interaction is due to the degeneracy of the conduction bands and the particular anisotropy used.

\subsection{Influence of the lattice degrees of freedom}
\label{sec:latt_degrees}
\begin{figure}[h]
\centering
\includegraphics[width=\linewidth]{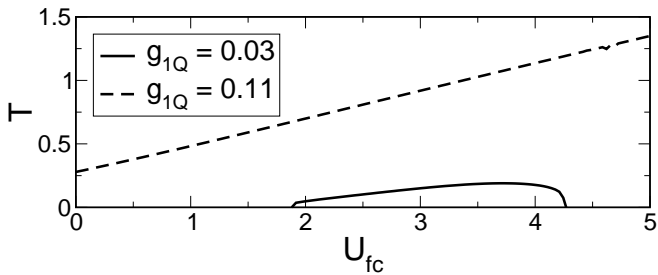}
\caption{CDW phase boundaries for $g_{1\Q}=0.03$ (solid line) and $g_{1\Q}=0.11$ (dashed line).}
\label{fig:critTemp}
\end{figure}

We now analyze the situation when phonons participate in the CDW formation. In Fig.~\ref{fig:critTemp} the critical temperatures for $g_{1\Q}=0.03$ and $g_{1\Q}=0.11$ can be found.
For very small electron-phonon coupling the phase diagram resembles the situation for the mEFKM.   
As the interaction strength  $g_{1\Q}$ increases the situation changes dramatically. For sufficiently large electron-phonon couplings, we 
no longer find critical lower and upper values $U_{fc}$ for the CDW transition and the transition temperature increases linearly with $U_{fc}$.  That is the critical temperature is significantly enhanced by $g_{1\Q}$. Evidently electron-hole attraction and  electron-phonon coupling act  together in the formation of a very stable CDW phase. 

\begin{figure}[h]
\centering
\includegraphics[width=\linewidth]{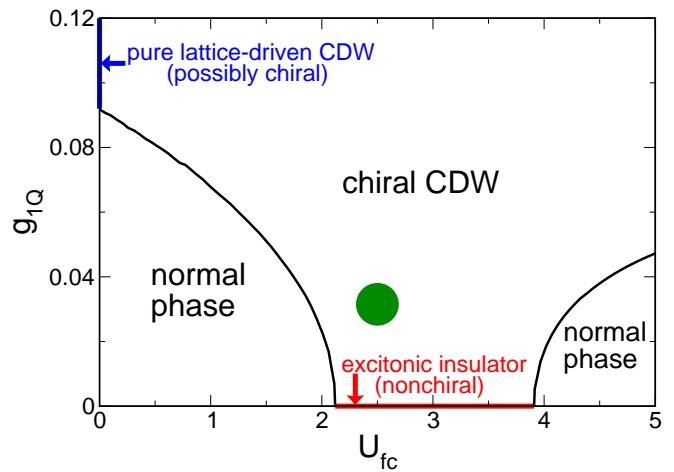}
\caption{(color online) Ground-state phase diagram of the mEFKM with additional electron-phonon coupling. 
The CDW phase is characterized by a finite gap parameter $|\tilde{\Delta}_\Q|$. The red line at $g_{1\Q}=0$ marks  the EI phase of the pure mEFKM. The blue line at $U_{fc}=0$ refers to a CDW induced solely by the electron-lattice interaction.  
The green point designates the range of model parameters appropriate for 1$T$-TiSe$_2$.}
\label{fig:GSPD}
\end{figure}

The impact of the Coulomb interaction and the electron-phonon interaction is summarized by the ground-state phase diagram shown in Fig.~\ref{fig:GSPD}. 
For weak electron-phonon couplings $g_{1\Q}$ the CDW is mainly driven by the Coulomb attraction $U_{fc}$ between electrons and holes. The greater $g_{1\Q}$, the larger the region where the CDW is stable. For $g_{1\Q} > 0.09$ the electron-phonon coupling alone can cause the CDW transition, even at $U_{fc}=0$  (blue line in Fig.~\ref{fig:GSPD}). Depending primarily on the magnitude of the static lattice distortion the CDW can be chiral in this limit, whereas the CDW  in the opposite  EI limit does not exhibit chirality  ($g_{1\Q}=0$, red line in Fig.~\ref{fig:GSPD}).

\subsection{Relation to 1$T$-TiSe$_2$}
\label{sec:TiSe2}
Based on the phase diagram derived for the mEFKM with electron-phonon coupling, we now attempt to estimate the electron-electron and electron-phonon interaction constants, $U_{fc}$ and $g_{1\Q}$, for 1$T$-TiSe$_2$. To make contact with experiments we take the displacements of the Ti ions measured by Di Salvo {\it et al.}: 
$\tilde u(n,m={\rm Ti})=0.04\mathring{A}$.\cite{SMW76} Then, from  Eq.~\eqref{latt_deform}, we can specify the value of $|\delta_\Q|$.  
For 1$T$-TiSe$_2$, the gap parameter was determined experimentally as $120$ meV by Monney {\it et al.}; see Ref.~\onlinecite{MSGMDBACMBBT10}.  Adjusting this value to our theoretical results yields $U_{fc}\approx2.5$ ($\approx 3$ eV) and $g_{1\Q}\approx0.03$ ($\approx 0.04$ eV); see the green marker in Fig.~\ref{fig:GSPD}. For these values both the theoretical ion displacement and gap parameter  are in the same order of  magnitude as the measured ones. Using $U_{fc}\approx2.5$ for 1$T$-TiSe$_2$, the electron-hole pairing is BCS-like. Since $g_{1\Q}\simeq 0.03$ is too small to cause a CDW for vanishing Coulomb interaction and, as discussed above, the EI scenario alone will not yield a stable chiral CDW, our results are in favor of a  combined lattice-deformation/EI mechanism for  the experimentally  observed chiral CDW transition, as suggested in Refs.~\onlinecite{WNS10, ZCS12}.

\section{Conclusions}
\label{sec:Summary}
In this work we have argued how the observed chiral charge-density-wave (CDW) phase in 1$T$-TiSe$_2$ may be stabilized. In the framework of the multiband extended Falicov-Kimball model (mEFKM) we showed that a purely electronic---exciton pairing and condensation---mechanism is insufficient to induce the observed  (long-ranged) chiral charge order. We propose that  the coupling of the electrons to the lattice degrees of freedom is essential for the formation of a chiral CDW state. 

We note that in our model clockwise and anticlockwise CDWs are degenerate. This is  in accord with experimental findings.\cite{ILSKITOT10} The chiral property can properly be observed in the ionic displacements accompanying the CDW in 1$T$-TiSe$_2$.

Whether the chiral CDW is stabilized depends particularly on the magnitude of the static lattice distortion and also on the ratios of the electron-phonon, respectively the phonon-phonon, interaction constants.
Our analysis confirms the sequential transition scenario $T_{\rm chiral\, CDW}<T_{\rm nonchiral\, CDW}$ as  was proposed in Refs.~\onlinecite{We11,We12} and corroborated experimentally.\cite{CROLGKRW12} However, we extended this scenario by the inclusion of further interactions. This leads to a CDW state for $T_{\rm chiral\, CDW}<T<T_{\rm nonchiral\, CDW}$, where the mirror symmetry is reduced compared to the normal phase, but chirality is not yet formed.

Concerning the microscopic mechanism underlying the CDW transition, we demonstrated that electron-electron interaction and electron-phonon coupling support each other in driving the electron-hole pairing and finally the instability.  This suggests that the CDW transition in 1$T$-TiSe$_2$ is due to  a combined lattice distortion and exciton-condensation effect. The outcome is a spontaneous broken-symmetry CDW low-temperature state with small but finite lattice deformation. Of course, both  the mean-field treatment of the Coulomb interaction and the frozen-phonon approach are rather crude approximations and a more elaborated study of the complex interplay between the electronic and phononic degrees of freedom  is highly desirable to confirm our proposed scenario for the chiral CDW transition in 1$T$-TiSe$_2$.

Let us finally point out that  we called $|\Delta_{\Q}|$ the excitonic-insulator order parameter on account of its analog in the EFKM.\cite{IPBBF08, SC08, ZIBF10, PBF10, ZIBF11, PFB11, ZIBF12, SEO11} The meaning of a finite $|\Delta_\Q|$ in the presence of a band coupling is  imprecise however. 
Likewise a spontaneous hybridization of the valence band with one of the conduction bands, signaling  the exciton condensate in the mEFKM,  
may  be induced by a sufficiently large electron-phonon coupling $g_{1\Q}$. A general criterion for the formation of an 
exciton condensate in a strongly coupled band situation has not been established to date.  This is an open issue which deserves further analysis because of its relevance in characterizing the nature of CDW transitions also in other materials.\cite{KTKO13,ELS12}

\begin{acknowledgements}
We thank P. Aebi, K. W. Becker, F. X. Bronold, D. Ihle,  G. Monney, and N. V. Phan for valuable discussions. This work is supported by the Deutsche Forschungsgemeinschaft through SFB 652 (project B5), by the Fonds National Suisse pour la Recherche Scientifique through Div. II, the Swiss National Center of Competence in Research MaNEP, and the U.S. Department of Energy. C.M. acknowledges also support by the Fonds National Suisse pour la Recherche Scientifique under grant PA00P2-142054.
\end{acknowledgements}

\bibliographystyle{apsrev4-1}
\bibliography{ref} 
\end{document}